\newcommand{\be}{\begin{equation}}
\newcommand{\ee}{\end{equation}}
\newcommand{\bea}{\begin{eqnarray}}
\newcommand{\eea}{\end{eqnarray}}
\newcommand{\ben}{\begin{eqnarray*}}
\newcommand{\een}{\end{eqnarray*}}

\newcommand{\newc}[1]{\textcolor{red}{#1}}

\documentclass[epj]{svjour}
\usepackage{amssymb}
\usepackage{amsmath}
\usepackage{graphicx}
\usepackage{amssymb}
\usepackage{epsfig}
\usepackage[dvips]{color}
\begin{document}
\title{The phase diagram of the
extended anisotropic ferromagnetic-antiferromagnetic Heisenberg chain}
\author{Evgeny Plekhanov\inst{1,2}, Adolfo Avella\inst{1,2,3}, and Ferdinando Mancini\inst{1,2}}
\institute{Dipartimento di Fisica ``E.R. Caianiello'',
Universit\`{a} degli Studi di Salerno, I-84084 Fisciano (SA), Italy\and
Unit\`{a} CNISM di Salerno, 
Universit\`{a} degli Studi di Salerno, I-84084 Fisciano (SA),
Italy\and
Laboratorio Regionale SuperMat, CNR-INFM, I-84084 Fisciano (SA), Italy}

\mail{plekhanoff@physics.unisa.it}

\authorrunning{E. Plekhanov \and A. Avella \and F. Mancini}

\titlerunning{The phase diagram of the extended anisotropic F-AF Heisenberg chain}
\date{\today}
\abstract{
By using Density Matrix Renormalization Group (DMRG) technique we study
the phase diagram of 1D extended anisotropic Heisenberg model with
ferromagnetic nearest-neighbor and antiferromagnetic
next-nearest-neighbor interactions. We analyze the static correlation
functions for the spin operators both in- and out-of-plane and classify
the zero-temperature phases by the range of their correlations. On
clusters of $64,100,200,300$ sites with open boundary conditions we
isolate the boundary effects and make finite-size scaling of our
results. Apart from the ferromagnetic phase, we identify two gapless
spin-fluid phases and two ones with massive excitations. Based on
our phase diagram and on estimates for the coupling constants known from
literature, we classify the ground states of several edge-sharing
materials.
}
%
%
%
\maketitle
\section{Introduction}\label{intro}
Recently, an increasing attention has been paid to the materials
containing edge-sharing ${\rm CuO_2}$ chains~({\textit e.g.}\\ ${\rm
Rb_2Cu_2Mo_3O_{12}}$ as in Ref.~\cite{hase}, ${\rm NaCu_2O_2}$ as
in Ref.~\cite{drechsler_01} or ${\rm LiCuVO_4}$ as in
Ref.~\cite{enderle_00}). It has been found by using several
complementary experimental techniques that the low-energy physics in
such materials is one-dimensional~\cite{enderle_00}. It has been also
concluded that in such insulating materials the spins, localized on the
copper ions, interact via ferromagnetic interaction with their nearest
neighbors, while a considerable next-nearest neighbor (NNN) interaction
has been argued to be antiferromagnetic. In addition, at least in one
material (${\rm LiCuVO_4}$), about 6$\%$ exchange anisotropy has been
measured by using paramagnetic resonance~\cite{nidda_00,vasiliev_00}. Another
manifestation of such anisotropy is the dependence of the saturation
value of the external magnetic field on its direction~\cite{enderle_00}.

It is widely accepted that to study such systems, a 1D extended
anisotropic Heisenberg model with ferromagnetic (F) nearest-neighbor
(NN) interaction and antiferromagnetic (AF) NNN one should be used. In
order to estimate the values of the exchange interactions in these
materials, temperature and magnetic field dependencies of the integrated
quantities, such as susceptibility and magnetization, have been compared
with those of various 1D spin models, calculated exactly on small
clusters. The F-AF Heisenberg model appeared to be the only
compatible~\cite{drechsler_01}. Excluding the symmetry breaking in
$XY$-plane, the resulting Hamiltonian could have at most four
parameters: two in-plane interaction constants and two out-of-plane
ones. Let us denote these as follows $J_{\perp}$, $J^{\prime}_{\perp}$,
$J_z$ and $J^{\prime}_z$.
Since one of them can always be used to set the unit of energy, there
are in fact only three independent interaction constants.
Such large number of parameters makes it
difficult to explore the full phase diagram, which would be
three-dimensional. There exist several parametrizations of this system,
which use less parameters. For instance, one can restrict oneself only
to isotropic case ($J_z=J_{\perp}$, $J^{\prime}_z=J^{\prime}_{\perp}$),
ending up with 
only one parameter {\it i.e.} $J_z/J_z^{\prime}$, and two cases,
according to the sign of $J_z^{\prime}$, as in the Ref.~\cite{tonegawa}.
Contrarily, one could allow for the same level of anisotropy both in NN
and NNN channels, which results in two parameters {\it i.e.} $J_{z}=\pm
1$, $J^{\prime}_z/J_{z}$ and $J_{\perp}/J_z\equiv
J^{\prime}_{\perp}/J^{\prime}_z$, as in the Ref.~\cite{hirata}. Yet,
another parametrization, adopted in the present paper, consists in
letting the anisotropy only in the NN channel, while leaving the NNN
interaction isotropic, as in Ref.~\cite{haldane_03}. This
parametrization amounts to have two parameters: $J_z=\pm 1$,
$J_{\perp}/J_z$ and $J^{\prime}/J_z\equiv
J^{\prime}_z/J_z=J^{\prime}_{\perp}/J_z$. 
Accordingly, the 1D extended
anisotropic Heisenberg model reads as follows (contrarily to
Ref.~\cite{haldane_03} we invert the sign of $J_z$):
\bea
   \nonumber
   H &=& -J_z \sum_i S^z_i S^z_{i+1} + J_{\bot}
   \sum_i ( S^x_i S^x_{i+1} + S^y_i S^y_{i+1})\\
   &+& J^{\prime}\sum_i \mathbf{S}_i \mathbf{S}_{i+2}.
   \label{ham}
\eea

\begin{figure}
   \includegraphics[angle=0, width=0.45\textwidth]{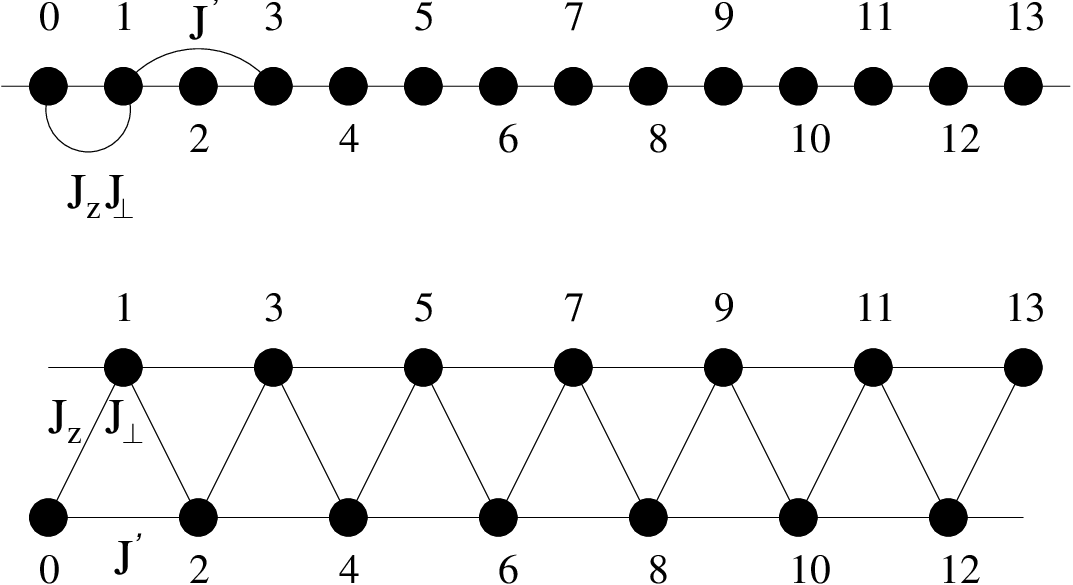}
   \caption{Mapping of the model~(\ref{ham}) on a two-leg zigzag ladder.}
   \label{fig_m}
\end{figure}

It is worth noting that the Hamiltonian~(\ref{ham}) with $J_{\perp}>0$
can be easily mapped onto the one with $J_{\perp}<0$. 
Indeed, for a bipartite lattice, the transformation:
\be
   \tilde{S}^{\alpha}_i = (-1)^i S^{\alpha}_i, \; \mathrm{where}\; \alpha=x,y.
\ee
inverts the sign in front of $J_{\perp}$ and changes an in-plane
correlation function $\langle S^{\alpha}_{i}S^{\alpha}_{i+n}\rangle$ by
the pre-factor $(-1)^{n}$. We choose AF sign of $J_{\perp}$ to be
compatible with our previous work~\cite{Plekhanov_01}, although in the
edge-sharing ${\rm CuO_2}$ materials $J_{\perp}$ should be negative.
While in edge-sharing materials the anisotropy has been
found at least in one compound~\cite{nidda_00,vasiliev_00}, its precise
structure and strength is still difficult to estimate. In this article, we intend to
study the qualitative effects of anisotropy and do not want to increase
excessively the number of parameters. That is why we
consider the anisotropy only in NN channel.
Finally, it is worth noting that the model~(\ref{ham}) can be also
represented as a two-leg zigzag ladder as show in Fig.~\ref{fig_m}.

From the theoretical point of view, F-AF Heisenberg model is a challenge
both for analytical and numerical methods. The model~(\ref{ham}) has
been extensively studied in its antiferromagnetic region~\cite{white_03}
($J_z<0$). For what concerns the ferromagnetic region ($J_z>0$), a few
results should be mentioned. First of all, it is worth noting that
despite the non-integrability of the model, there exist two points in
its parameter space where the analytic expression for the ground state
energy and wave functions are known. These are the points: ($J_z=-1$,
$J_{\perp}=J_z$, $J^{\prime}= J_{z}/2$) as proven in
Ref.~\cite{majumdar} and ($J_z=1$, $J_{\perp}=J_z$, $J^{\prime}=
J_{z}/4$) as found in Ref.~\cite{hamada}. Secondly, the famous Haldane
conjecture~\cite{haldane_01,haldane_02} states that one-dimensional AF
NN spin chains, composed of half-integer spins, could only have massless
excitations and power-law-decaying correlation functions. Another
conjecture -- the so-called ``ladder conjecture'' -- comes from
numerical methods~\cite{dagotto,hung},
bosonization~\cite{affleck_01,schultz} and subsequently from experiments
on a series of ladder materials like Sr$_{(n-1)/2}$Cu$_{(n+1)/2}$O$_n$
$(n=3,5\ldots)$~\cite{azuma,kojima}. This conjecture states:
``spin-$\frac{1}{2}$ ladders composed of an even number of chains have
gapped excitations, while those with an odd number of chains have
gapless excitations''. 
However, none of these arguments carries over to the extended
anisotropic Heisenberg model~(\ref{ham}) and therefore the issue of
whether it could possess a gapped ground state and whether the gap can
be observed numerically is still controversial.
The existence of an astronomically small gap~\cite{itoi} (with a
correlation length of the order of $10^{36}$ lattice spacings) has been
predicted by means of the Renormalization Group analysis of the
effective field theory in the limit of two AF chains coupled by a weak F
inter-chain coupling. A gapped dimer phase, surrounded by the gapless
spin-fluid ones, has also been identified by using the level-crossing
analysis of the excited states obtained by Lanczos diagonalization on
small rings~\cite{nomura,somma_01}. Similar results were found by using
the Quantum Renormalization Group~\cite{jafari}. In addition, a
dimerized gapped phase~\cite{bursill}, incommensurate power-law
correlations~\cite{nersesyan} and chiral
order~\cite{kaburagi_00,kaburagi_01,kaburagi_02} have been found on the
AF side of the model ($J_z<0$). On the other hand, in
Ref.\cite{dmitriev_01} by means of the perturbation theory around the
critical point ($J^{\prime}=0.25 J_z$, $J_{\perp}=J_z$) of weakly
anisotropic F-AF Heisenberg model, only spin-fluid gapless phases have
been found. This result might be an artefact of the perturbation theory,
since a perturbation can not change the underlying mean-field ground
state unless one is summing up an infinite number of diagrams. Finite
magnetic field phase diagram of the model~(\ref{ham}) has been studied
in the Refs.~\cite{hm_00,hm_01,hm_02,sudan_00}, which gave an insight
into the nature of elementary excitations unveiling both gapped and
gapless excitations depending on the values of Hamiltonian parameters.

Even for the NN Heisenberg model~($J^{\prime}=0$),
despite the existence of exact Bethe-ansatz solution, it is still
extremely difficult to obtain closed analytic expressions for the
correlation functions (see {\it e.g.}~\cite{kitanine_01,kitanine_02}). 
Asymptotic analytical results
come mainly from the Quantum Field Theory. In Ref.~\cite{affleck}
it has been shown, based on the Conformal Field Theory, that the
long-range behavior ($r\to\infty$) for the isotropic antiferromagnetic
Heisenberg model has a highly non-trivial form:
\be
   \langle S^{\alpha}_0 S^{\beta}_r \rangle \to \delta_{\alpha,\beta}
   \frac{(-1)^r}{(2\pi)^{3/2}}
   \frac{\sqrt{(\ln r)}}{r},
\ee
where $\alpha$, $\beta=x,y,z$. As soon as the rotational invariance is
broken, however, the correlation functions of the nearest-neighbors
anisotropic Heisenberg (XXZ) model assume a simple power-law
form:
\bea
    \label{xy-1}
    \langle S^{x}_0 S^{x}_r \rangle &\to& (-1)^r A_x r^{\eta} + B_x r^{\eta + 1/\eta}\\
    \label{xy-2}
    \langle S^{z}_0 S^{z}_r \rangle &\to& (-1)^r A_z r^{1/\eta}   + B_z r^{-2}.
\eea
Here $\langle S^{y}_0 S^{y}_{r\phantom{0}}\rangle=\langle S^{x}_0
S^{x}_r\rangle$, $\eta=-\arccos(J_z/J_{\perp})/\pi$, $0\le
|J_z/J_{\perp}|<1$, $A_x$ has been determined in
Ref.~\cite{lukyanov} and $A_z, B_x, B_z$ are not known in general
for arbitrary values of the ratio $J_z/J_{\perp}$.

\begin{figure*}
\begin{center}
   \includegraphics[angle=270, width=0.49\textwidth]{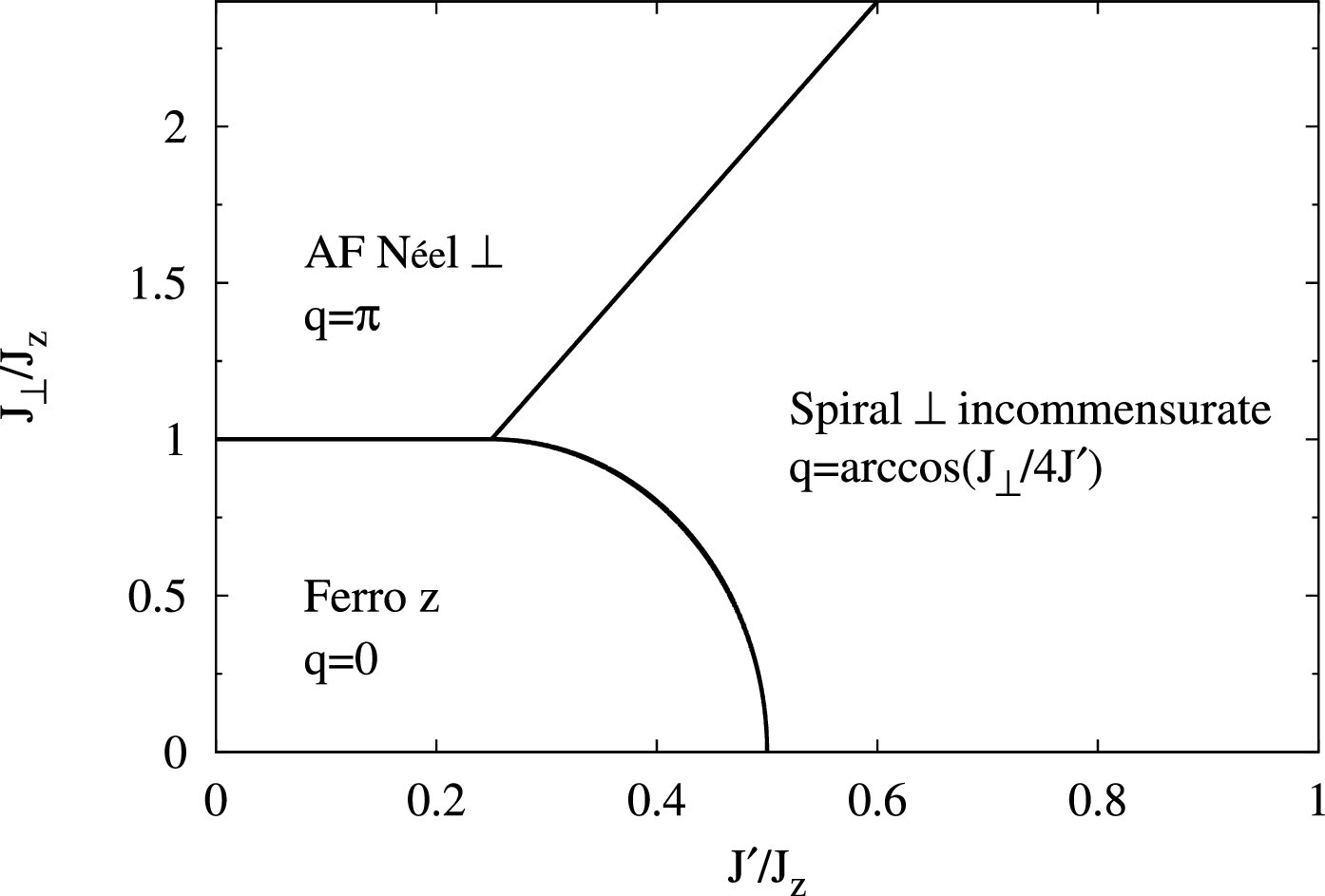}
   \includegraphics[angle=270, width=0.49\textwidth]{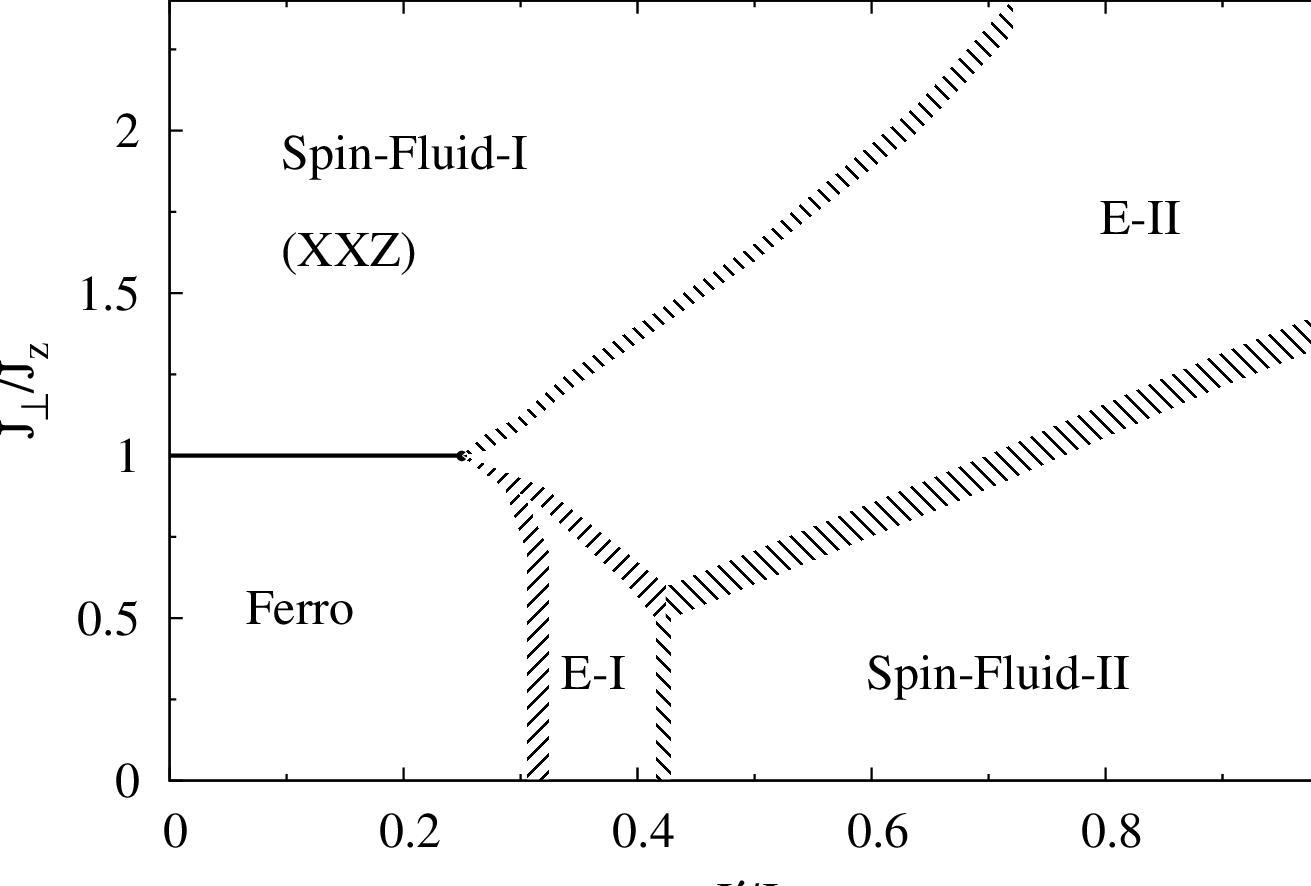}
   \caption{
   Quantum phase diagram of the Hamiltonian~(\ref{ham}) (right
   panel) as compared to the classical one (left panel).
   The dashed regions are the phase boundaries determined with
   systematic error owing to the finiteness of the cluster size.
   }
   \label{fig_phd}
\end{center}
\end{figure*}

\begin{figure*}
   \begin{center}
	  \includegraphics[angle=0,width=1.005\textwidth]{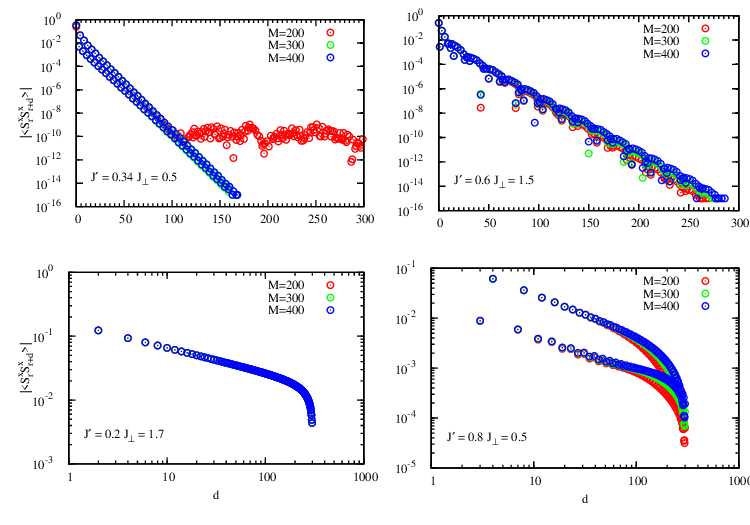}
	  \caption{
	  A study of DMRG truncation error as a function of the number of
	  retained density matrix eigenstates $M$ at four points representing
	  the phases E-I(top-left panel), E-II(top-right),
	  Spin-Fluid-I(bottom-left), Spin-Fluid-II(bottom-right). In-plane
	  correlation function is taken as an example.
	  }
	  \label{figM}
   \end{center}
\end{figure*}
In the past we have already considered the ferromagnetic ($J_z>0$)
model~(\ref{ham}) in connection with the question of ergodicity of the
system's dynamics~\cite{Plekhanov_01,Plekhanov_02}, by
means of the Lanczos and Exact Diagonalization techniques. Therein a
zero-temperature phase diagram has been constructed, based on whether
the dynamics of the $z$-projection of local spin was ergodic or not.
Different types of ergodic phases have been identified depending on how
the non-ergodic constant was approaching zero within the finite-size
scaling. Another insight into the physics of~(\ref{ham}) has been made
from the entanglement studies of the phase transitions in the vicinity
of the ferromagnetic phase. Two different types of behavior can be identified,
depending on whether $J^{\prime}$ or $J_{\perp}$ is 
increased~\cite{Plekhanov_04}.
By means of Lanczos, it was impossible to understand deeper the nature of
the underlying phases without analyzing the asymptotic behavior of the
correlations at large distances and accurate finite-size scaling for
large clusters. That is why we revisit the phase diagram of~(\ref{ham})
in the present manuscript using the Density Matrix Renormalization Group
(DMRG).

This article is organized as follows: after the definition of the model
and the method used to analyze it in Section~\ref{model}, we
briefly revisit the classical phase diagram of~(\ref{ham}) at $T=0$ in
Section~\ref{class}. In Section~\ref{gen}, we describe the various phases
found in the quantum phase diagram of~(\ref{ham}).
\section{Method of analysis} \label{model}
In the present work, we focus our attention on studying the
spin-$\frac{1}{2}$ one-dimensional anisotropic Heisenberg model
with NNN interaction~(\ref{ham}). We use $J_z>0$, which corresponds to
ferromagnetic coupling.

We obtain the ground state properties of the Hamiltonian (\ref{ham})
numerically by means of the DMRG~\cite{white_01,white_02} technique on
chains with $L=64,100,200,300$ sites, subject to open boundary
conditions (OBC). In the DMRG calculations, we maintain up to $M=200$ lowest
eigenstates of the reduced density matrix in the basis of each DMRG
block, which permitted us to obtain a truncation error
on the sum of retained density matrix eigenvalues of the order of
$10^{-6}$. The real-space spin-spin correlation functions are calculated by
means of the finite-system algorithm.
We have also studied the systematic error caused by the truncation of
the density matrix eigenvalue basis. For every phase found in the phase
diagram of~(\ref{ham}), we have taken a representative point and
calculated, as an example, the in-plane correlation function for
$M=200,300,400$ as shown in Fig.\ref{figM}. It can be seen therefrom
that for the phases Spin-Fluid-I and E-II there is no significant
improvement by increasing $M$, while for the phase Spin-Fluid-II the
linearity of the correlation function extends towards the end-points of
the cluster. In the E-I phase, due to strong exponential decay of the
correlations, at lengths of the order of $100$ sites the absolute value
of the correlation function becomes less than $10^{-10}$ and goes beyond
the capabilities of DMRG at $M=200$. The increase of $M$ improves the
accuracy and the trend at small $d$ extends to a wider range of
distances. Anyhow, the improvement by increasing $M$ would lead to
negligible corrections to the slopes of the curves in Fig.\ref{figM},
{\it i.e.} critical exponents or correlation lengths.

A few words should be said about the determination of the phase
boundaries on a finite-size cluster. When the correlation length
becomes comparable to the cluster size, it is not possible to determine
the exact location and the order of the transition, but only the region
in the phase diagram, where the transition should occur in the bulk.
Such region usually shrinks increasing the cluster size. Moreover, the
properties of a cluster with OBC converge slower to the thermodynamic
limit in comparison to those of the clusters with {\it e.g.} periodic
boundary conditions. On the other hand, the method of
level-crossing~\cite{nomura,somma_01}, successfully applied to the
determination of the phase boundaries in Lanczos and Exact
Diagonalization techniques, cannot be used efficiently in DMRG
calculations, since clusters with OBC lack many of the important
symmetries. Nevertheless, DMRG has the important advantage of being able
to simulate systems as large as several hundreds of sites. In an OBC
system, the boundary effects penetrate inside the system at some finite
length $\lambda_b$ ($\lambda_b/a\lesssim20$ in our analysis, where $a$
is the lattice constant). Therefore, on a cluster of hundred or several
hundreds of sites, the central part of the cluster behaves effectively
as a bulk system. The long-range part of the correlations can be safely
observed in this part of the cluster. In the present paper, the
transitions we deal with are those between phases with at least one
massive mode and phases which seem not to have any.
In such a case, the phase boundary can be estimated
from the massive side as the point where the correlation length becomes
of the order of the cluster size, while from the massless side as the
point where the power-law behavior ceases to be observed. The difference
between these two points is a measure of uncertainty in determining the
exact position of the transition line. This uncertainty decreases with
the system size.
\section{Classical phase diagram}\label{class}
Before considering the phase diagram of~(\ref{ham}) it is instructive to
take a look at the classical limit of this model and its phase diagram at
$T=0$. The calculations are rather straightforward and will be just
sketched here. In classical case, the spins are represented by the
classical vectors of constant length $s$. One has to minimize the energy
functional of the system, containing the scalar products between the
nearest and next-nearest vectors, subject to the constraint that the
length of each vector be $s$. Fourier transform brings the Hamiltonian in the
"diagonal" form ({\it i.e.} the operators $S^{\alpha}_q$ and
$S^{\alpha}_{q^{\prime}}$, where $\alpha=x,y,z$, are not coupled unless
$q^{\prime}=-q$):
\bea
   \nonumber
   &&H= -J_z \sum_q \cos q   S^z_q S^z_{-q} 
   +J^\prime \sum_q \cos 2q \mathbf{S}_q \mathbf{S}_{-q}\\
   &&+J_\perp  \sum_q \cos q ( S^x_q S^x_{-q} + S^y_q S^y_{-q})=\\
   \nonumber
   &&\sum_q \gamma_z(q) S^z_q S^z_{-q} +
   \sum_q \gamma_{\perp}(q)( S^x_q S^x_{-q}
	+ S^y_q S^y_{-q}).
\eea
Here $\gamma_z(q)$ and $\gamma_{\perp}(q)$ are defined as follows:
\bea
   \gamma_z(q)      &=&-J_z \cos q + J^{\prime} \cos 2q\\
   \gamma_{\perp}(q)&=&\phantom{-} J_{\perp} \cos q + J^{\prime} \cos 2q.
\eea
In such circumstances it is easy to see that the global minimum
of the energy is realized by taking only one component with $q=q_0$ and
maximal value $S_{q_0}=s\sqrt{N}$ and putting all the others to zero.
$q_0$ is chosen such that either of the following conditions takes
place:
\bea
   \nonumber
   \gamma_z(q_0) &\to& \min; \quad \gamma_z(q_0) < \gamma_{\perp}(q),
   \forall q\\
   \nonumber
   \gamma_{\perp}(q_0) &\to& \min; \quad \gamma_{\perp}(q_0) < \gamma_{z}(q).
   \forall q
\eea
In the former case all spins are aligned along the $z$-axis, while in the
latter they are parallel to the $XY$-plane.
\begin{figure*}
   \includegraphics[angle=270, width=0.47\textwidth]{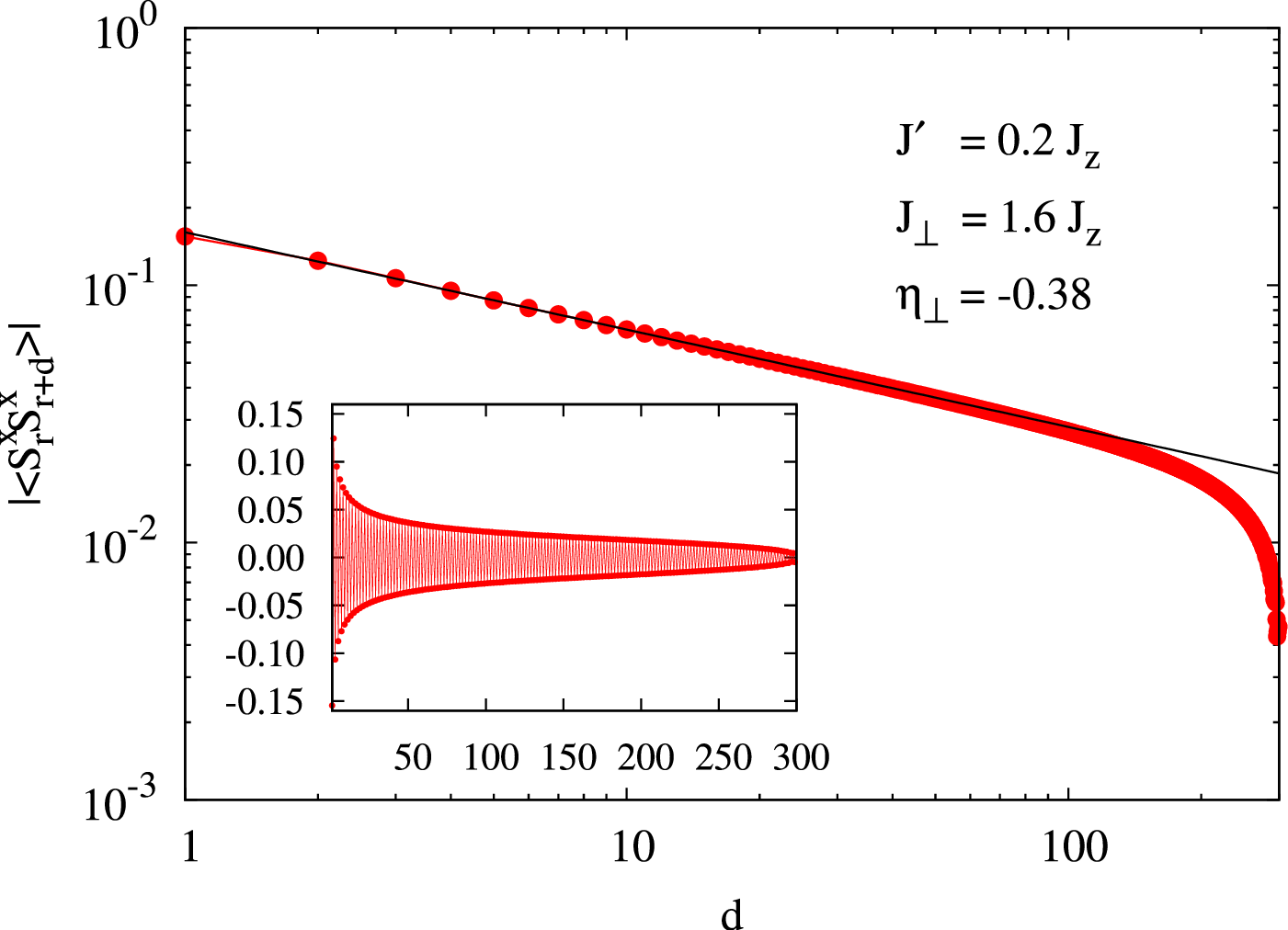}
   \hspace{0.4cm}
   \vspace{0.4cm}
   \includegraphics[angle=270, width=0.47\textwidth]{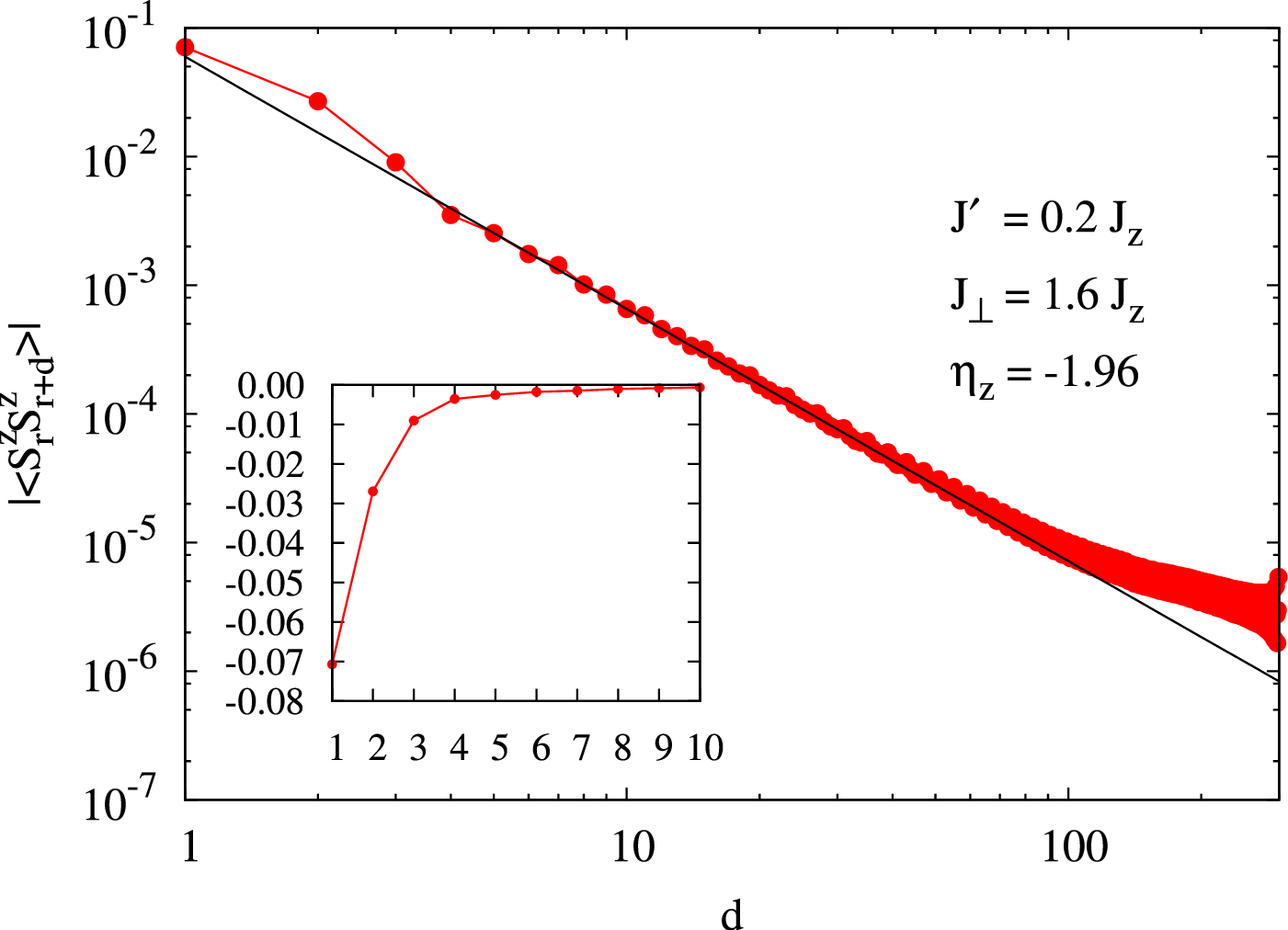}
   \caption{(Color online). $300$ site cluster, $J^{\prime}=0.2 J_z$,
   $J_{\perp}=1.6 J_z$. Example of static correlation functions for
   in-plane (left panel) and out-of-plane (right panel) channels plotted
   in logarithmic scale in the Spin-fluid-I (XXZ) phase. In the insets,
   the correlation functions are reported in linear scale.
   }
   \label{fig_xxz}
   \includegraphics[angle=270, width=0.47\textwidth]{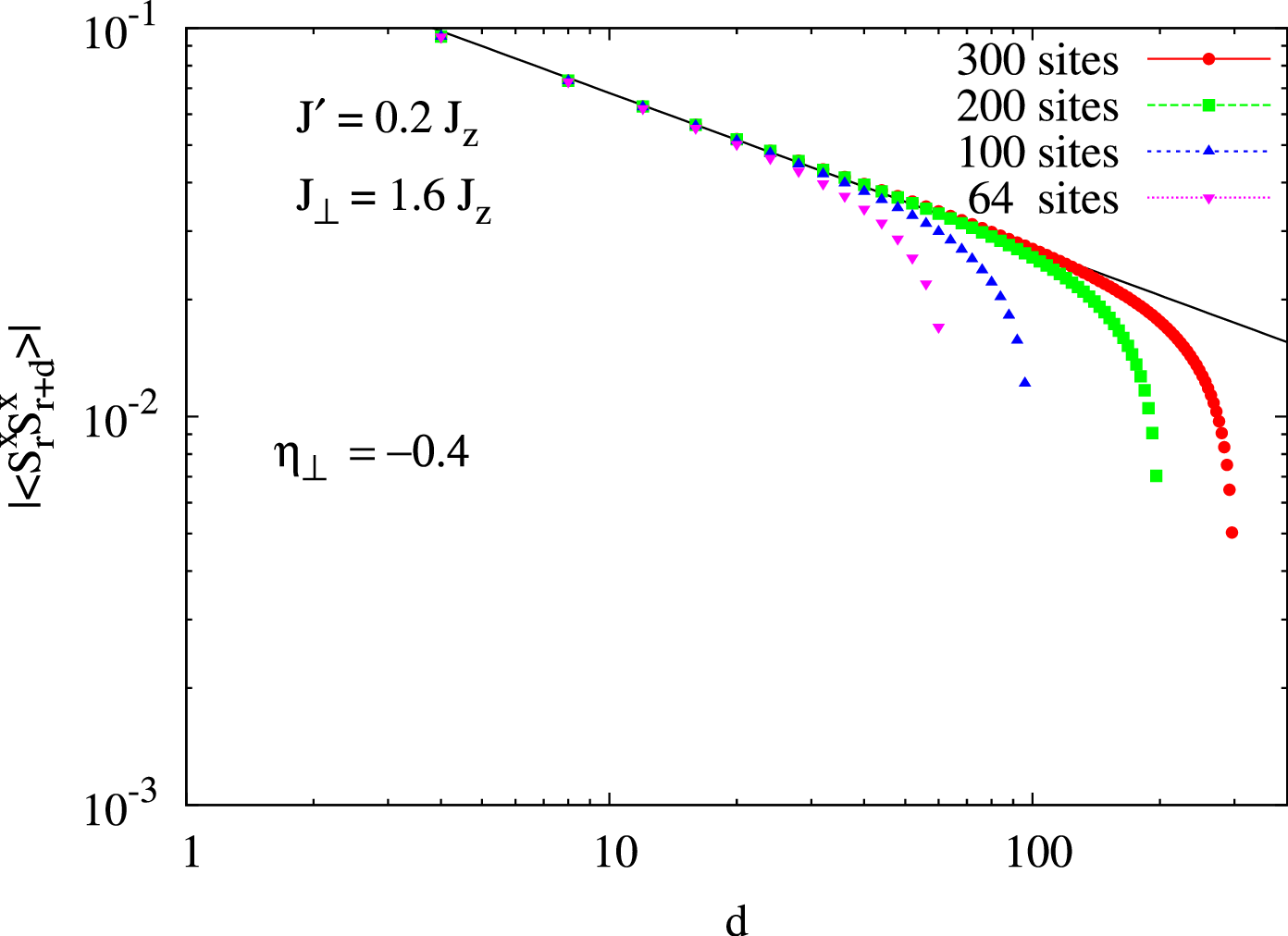}
   \hspace{0.4cm}
   \includegraphics[angle=270, width=0.47\textwidth]{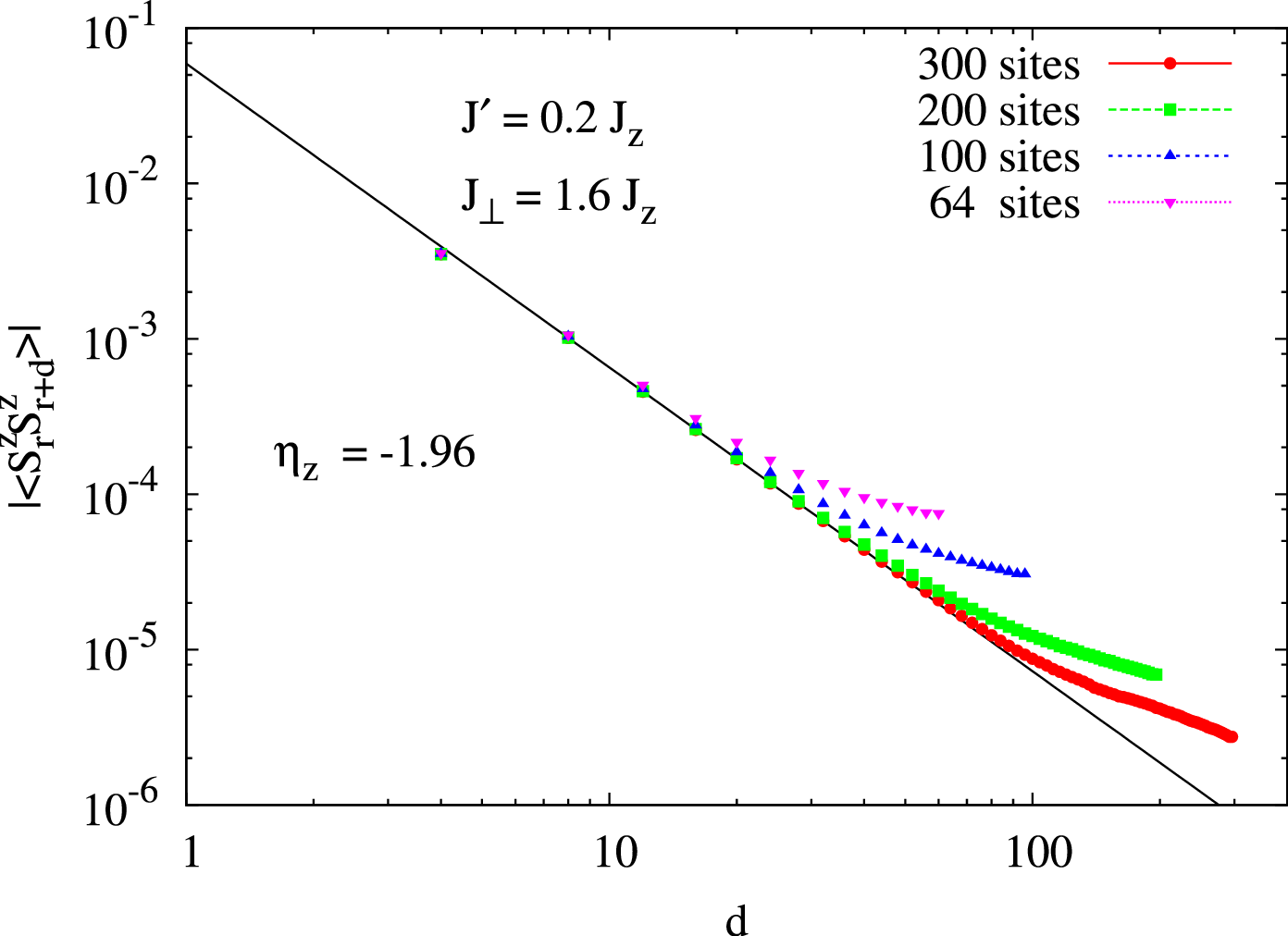}
   \caption{(Color online). Correlation functions in the Spin-Fluid-I
   phase at $J^{\prime}=0.2J_z$, $J_{\perp}=1.6J_z$ for different
   system sizes in the in-plane channel (left panel) and out-of-plane
   channel (right panel).}
   \label{fig_fss1}
   \includegraphics[angle=270, width=0.47\textwidth]{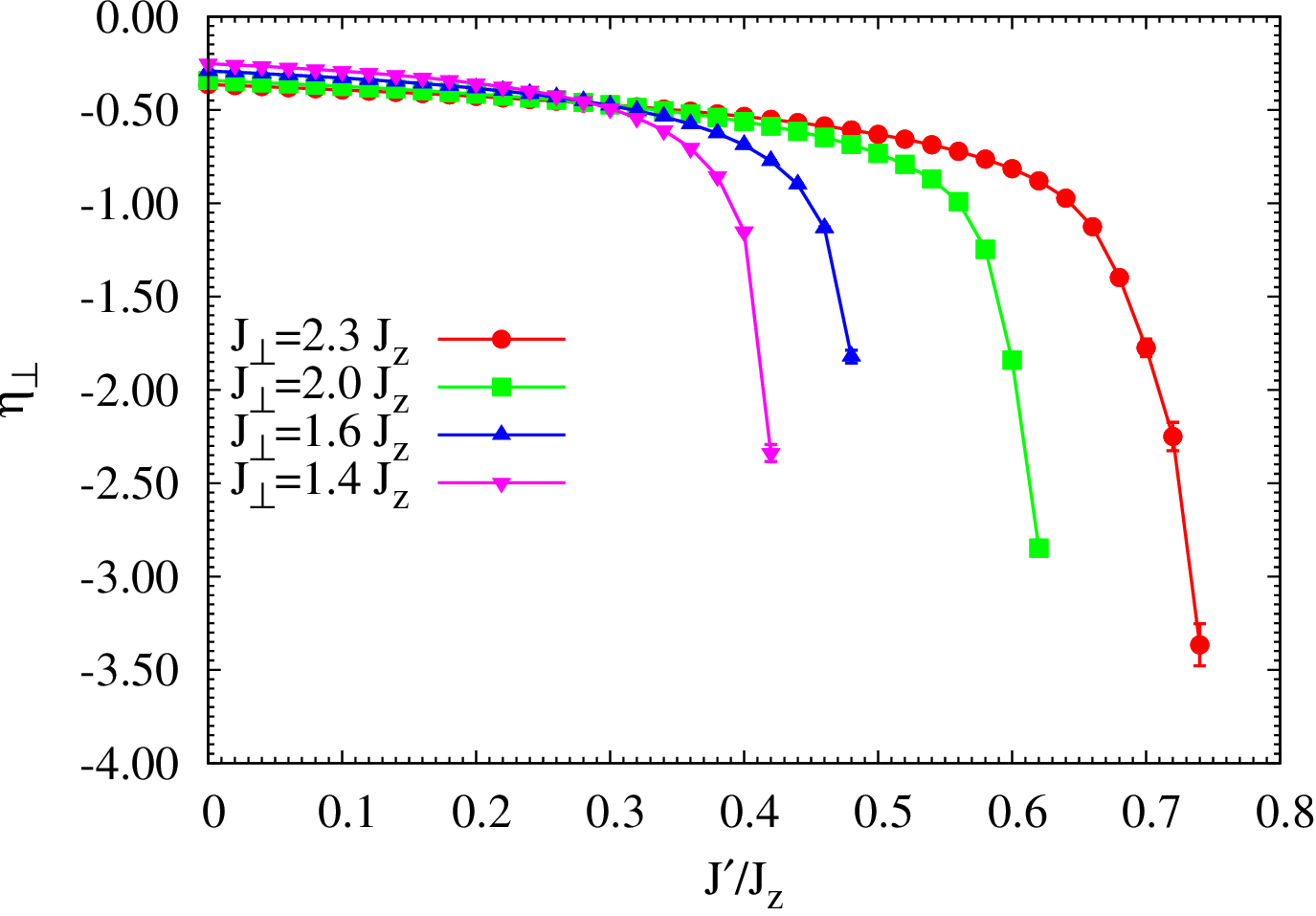}
   \hspace{0.4cm}
   \includegraphics[angle=270, width=0.47\textwidth]{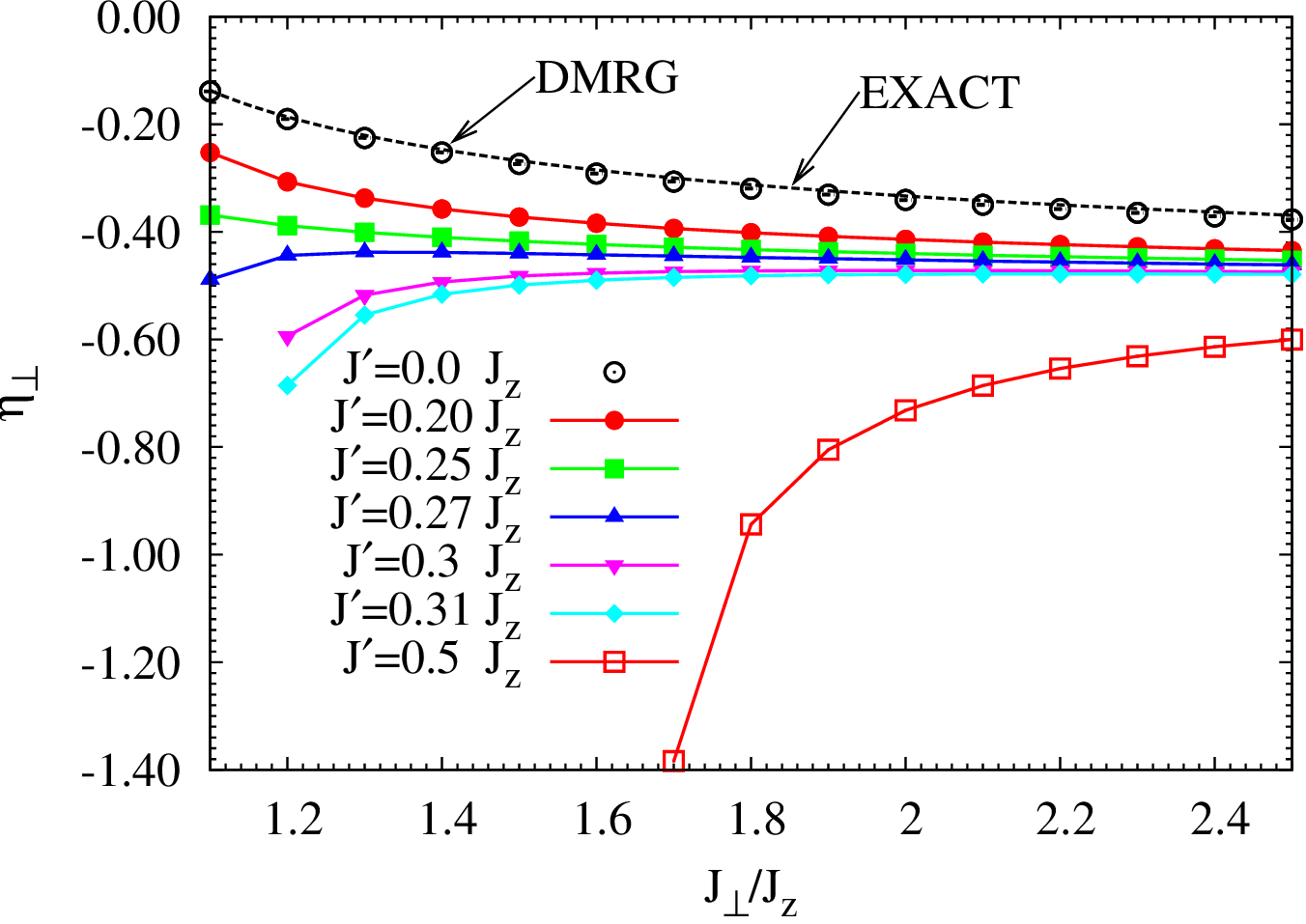}
   \caption{(Color online). $300$ site cluster. Spin-fluid-I phase:
   critical exponent of the in-plane correlations $\eta$ as a function
   of $J^{\prime}$ (left panel) and that of $J_{\perp}$ (right panel).
   The black curve labelled ``exact'' is the bosonization result from
   formula~(\ref{xy-1})}
   \label{fig_xx}
\end{figure*}
The above conditions give rise to the classical phase diagram depicted
in Fig.~\ref{fig_phd} (left panel). In the absence of quantum
correlations the system possesses a perfect long-range order, either
commensurate or incommensurate, with periodicity vectors $q$ as shown in
Fig.~\ref{fig_phd}. There are three phases in the classical limit of
the model~(\ref{ham}) in the range $J^{\prime}>0$, $J_{\perp}>0$. The
first one is the ferromagnetic phase (Ferro) with all spins fully
polarized along the $z$-axis. The second phase is an in-plane N\'eel
antiferromagnet, while the third one is the in-plane incommensurate
spirally ordered phase with the periodicity vector
$q=\arccos(J_{\perp}/4J^{\prime})$. The Ferro phase is separated from
the N\'eel one by the line $J_{\perp}=J_z$, $J^{\prime}<J_z/4$, while
the N\'eel phase is separated from the Spiral one by the line
$J_{\perp}=4J^{\prime}$, $J_{\perp}>J_z$. Finally, the Spiral phase is
separated from the Ferro one by the curve
$J_{\perp}=\sqrt{8J^{\prime}(1-2J^{\prime})}$, $J_z/4<J^{\prime}<J_z/2$.
\begin{figure*}
   \includegraphics[angle=270, width=0.47\textwidth]{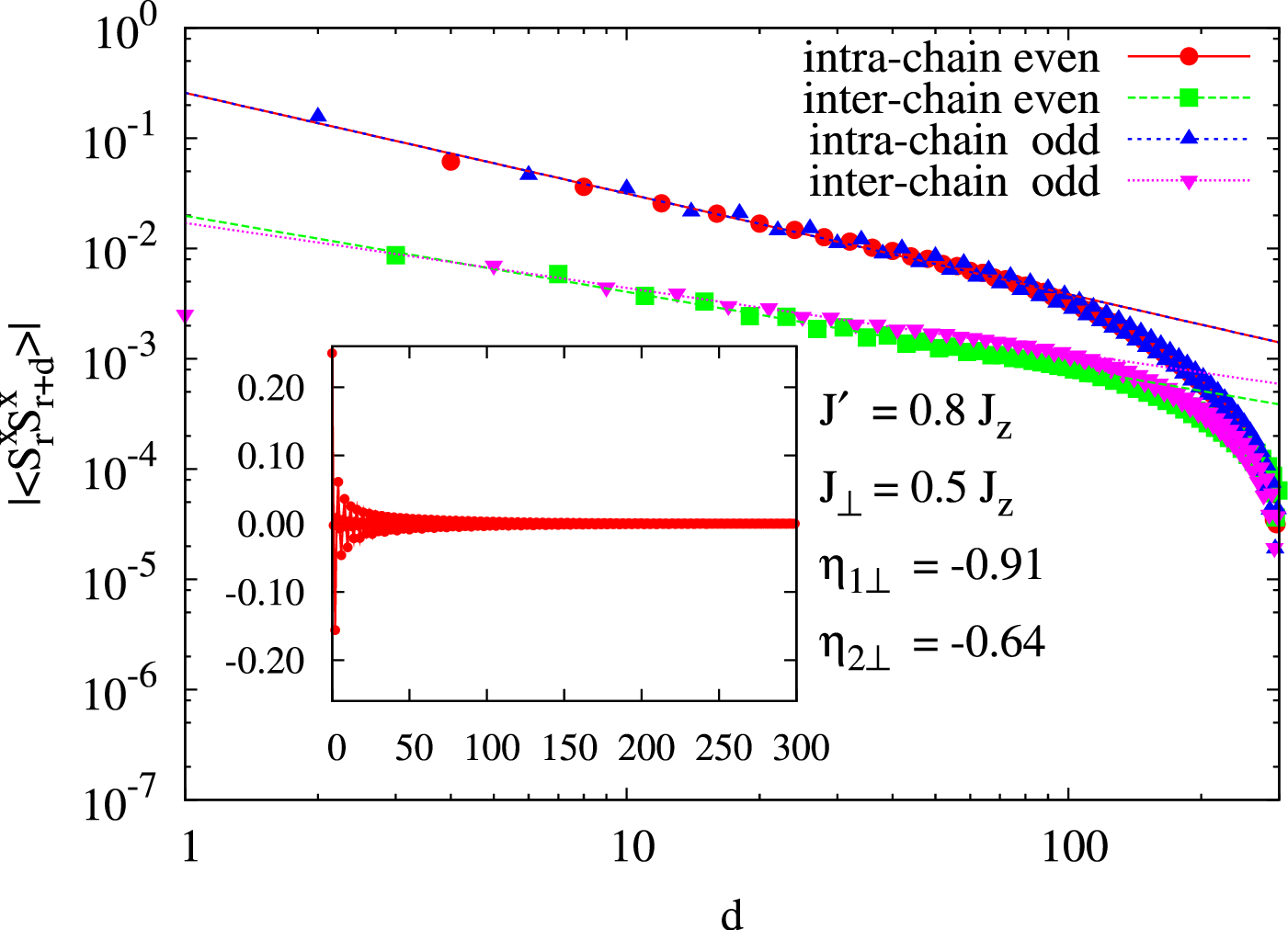}
   \hspace{0.4cm}
   \vspace{0.4cm}
   \includegraphics[angle=270, width=0.47\textwidth]{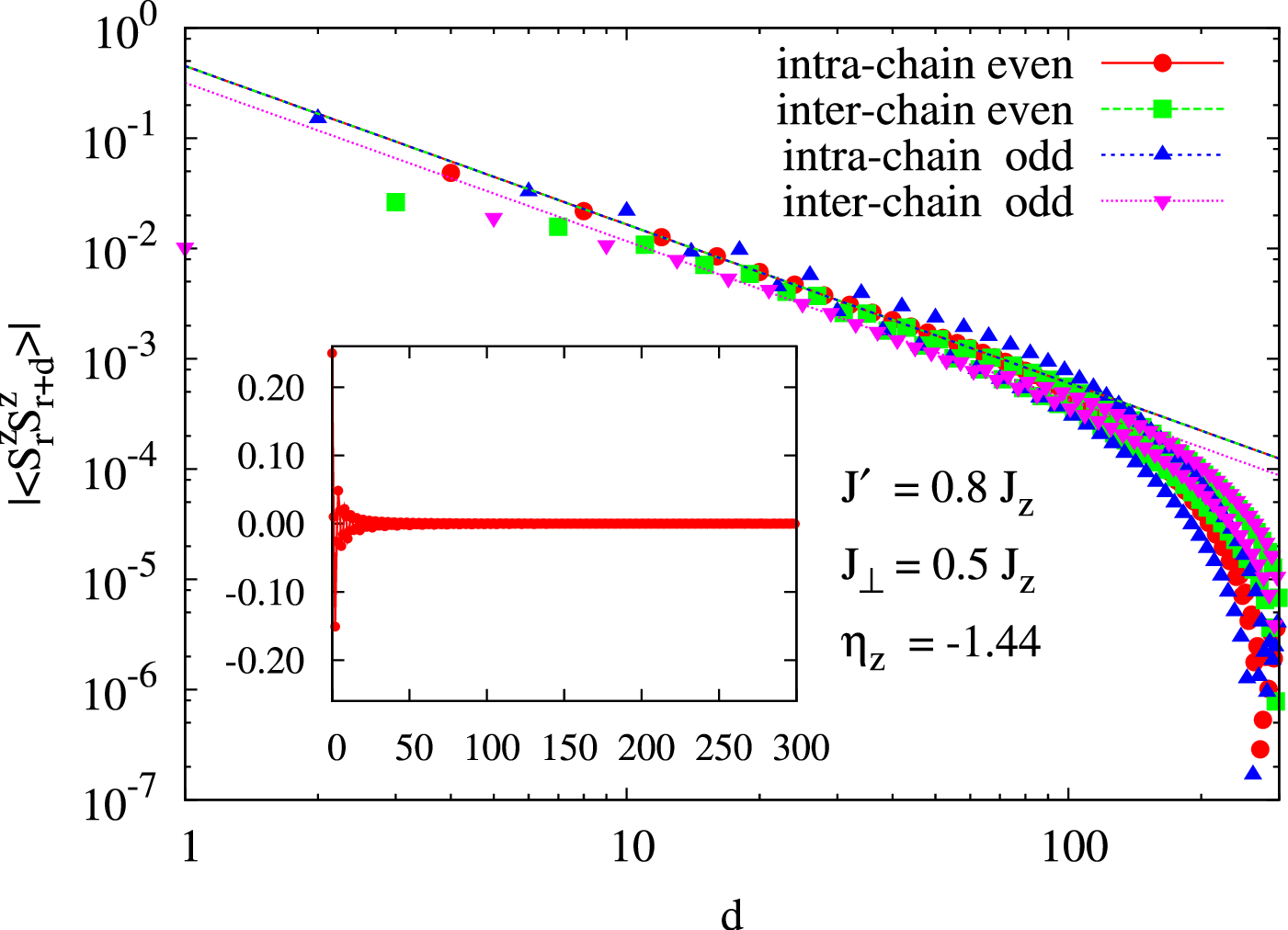}
   \caption{(Color online). $300$ site cluster, $J^{\prime}=0.8 J_z$
   $J_{\perp}=0.5 J_z$. Example of static correlation functions for
   in-plane (left panel) and out-of-plane (right panel) channels plotted in
   logarithmic scale in the Spin-fluid-II phase. In the insets,
   the correlation functions are reported in linear scale.
   }
   \label{fig_ii}
   \includegraphics[angle=270, width=0.47\textwidth]{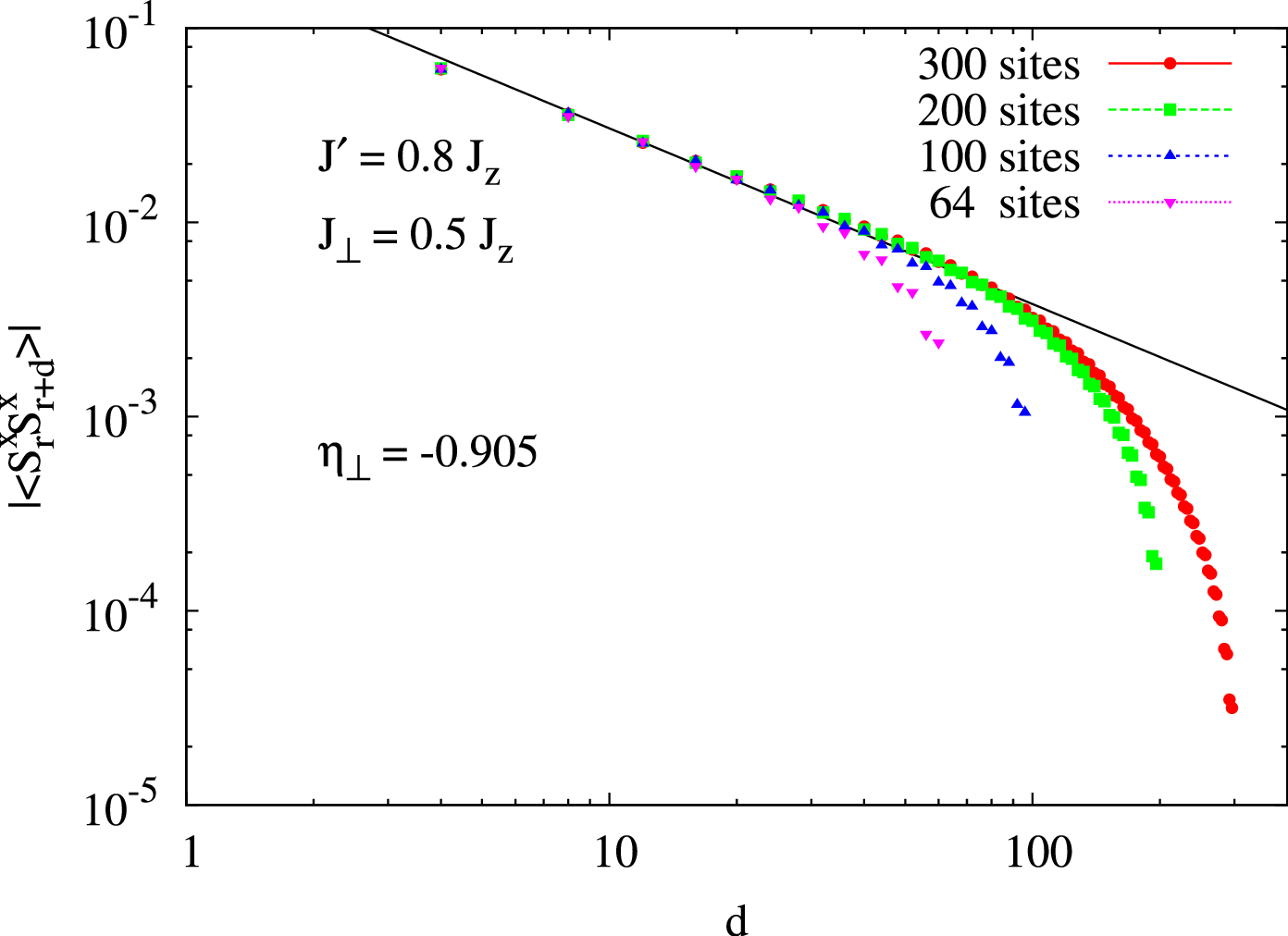}
   \hspace{0.4cm}
   \includegraphics[angle=270, width=0.47\textwidth]{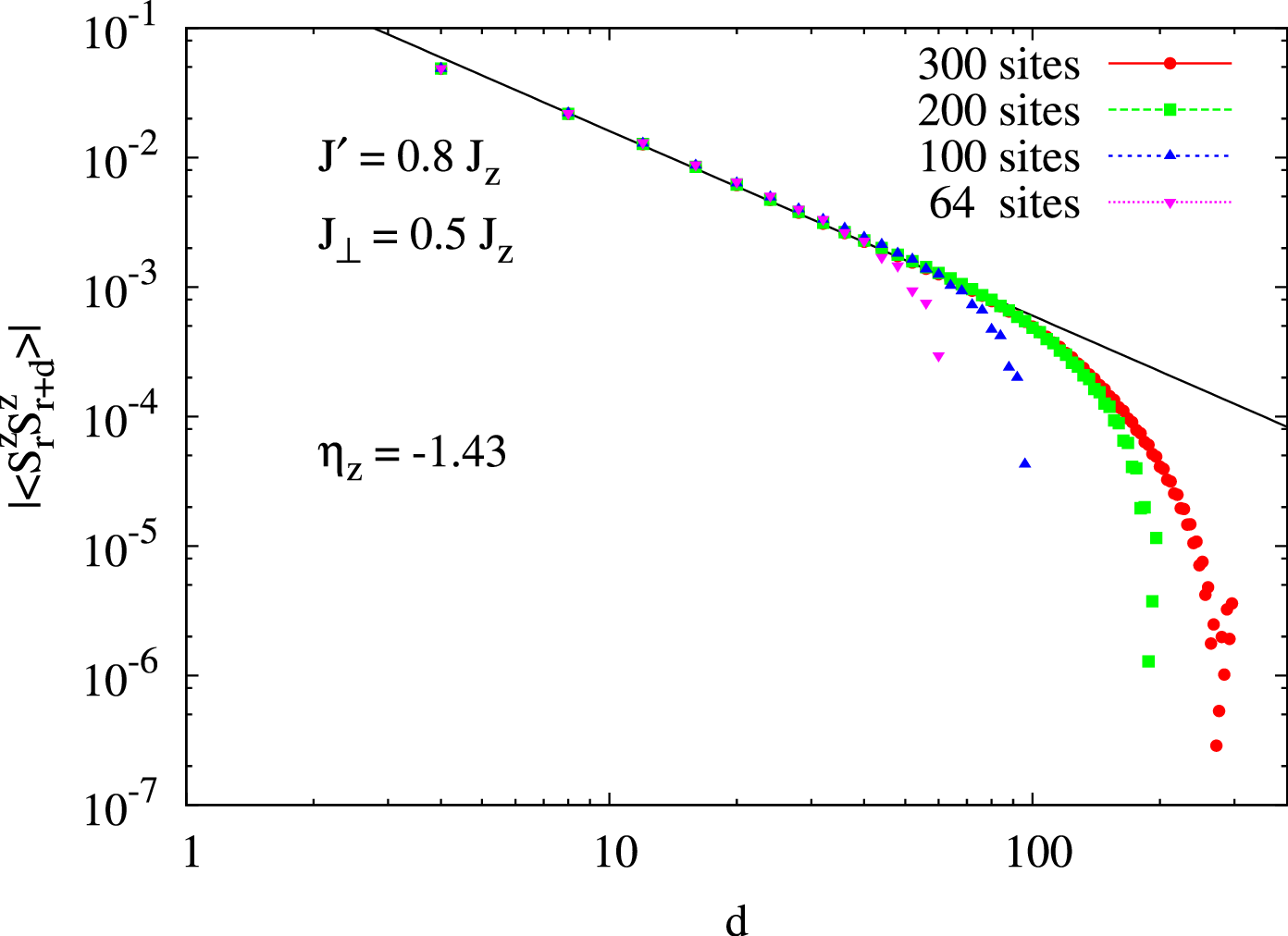}
   \caption{(Color online). Correlation lengths of the intra-chain
   even-distance correlation functions in Spin-Fluid-II phase at
   $J^{\prime}=0.5J_z$, $J_{\perp}=0.8J_z$ for different system sizes
   in the in-plane channel (left panel) and out-of-plane channel (right
   panel).
   }
   \label{fig_fss2}
   \includegraphics[angle=270, width=0.47\textwidth]{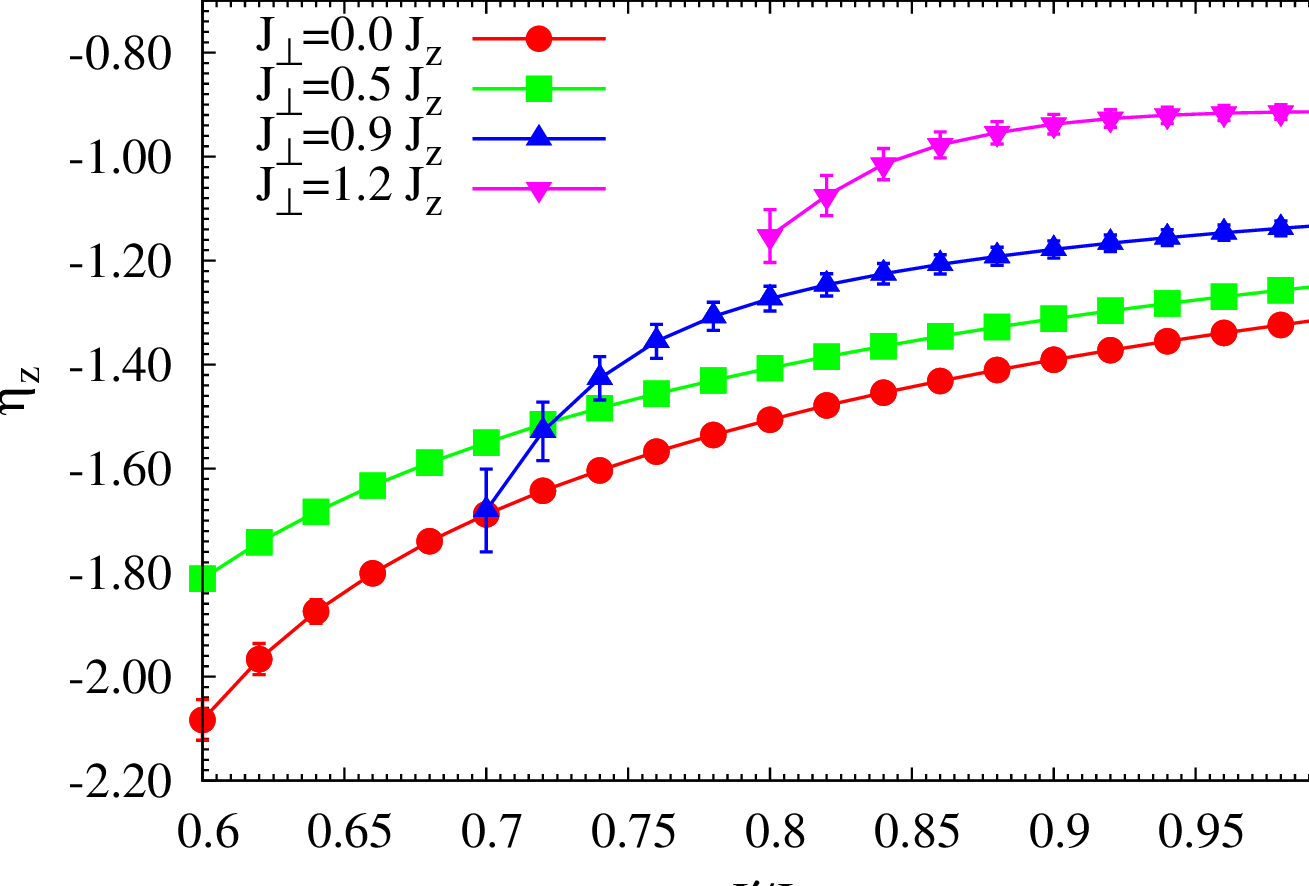}
   \hspace{0.4cm}
   \includegraphics[angle=270, width=0.47\textwidth]{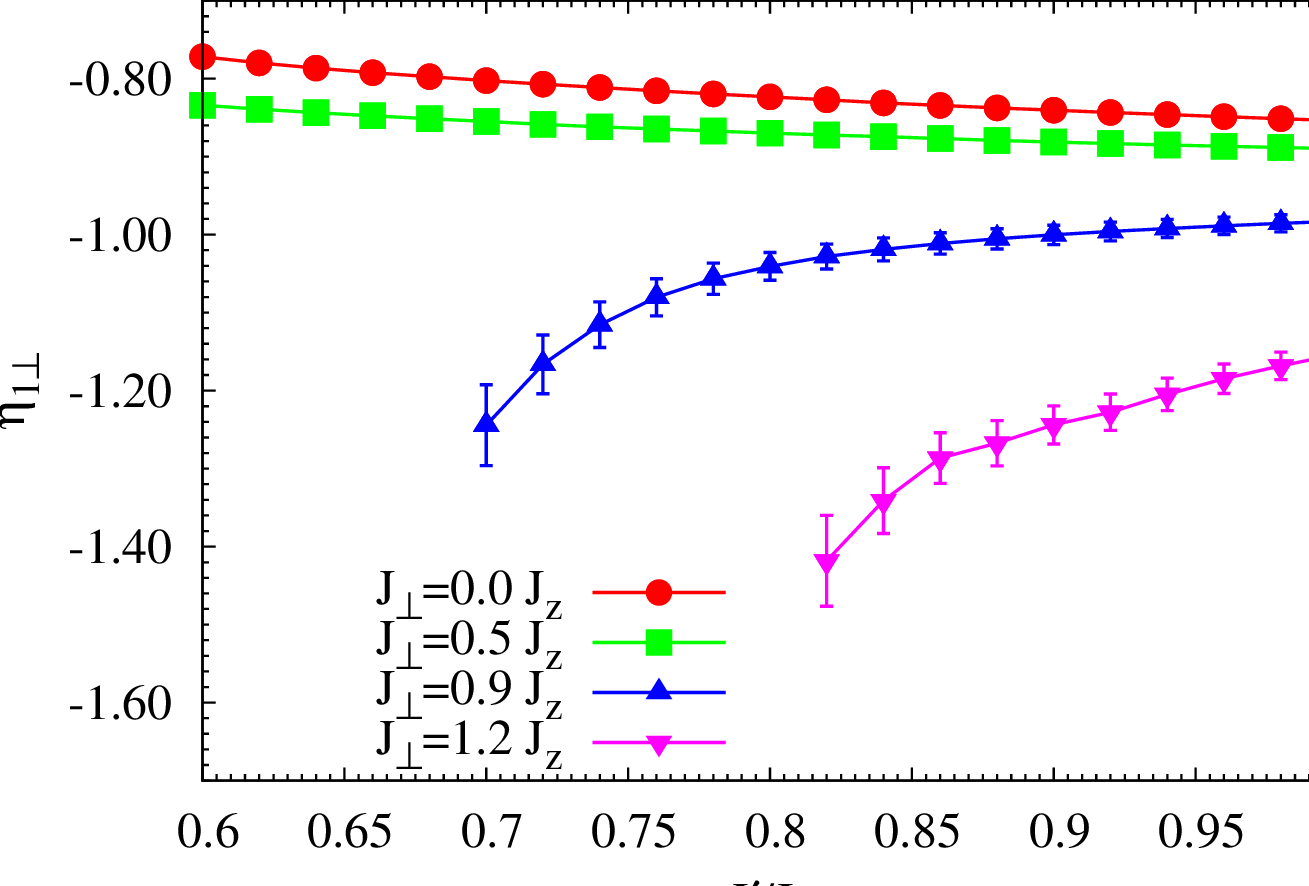}
   \caption{(Color online). $300$ site cluster. Spin-fluid-II phase:
   intra-chain critical exponent of the out-of-plane correlations
   $\eta_z$ (left panel) and that of the in-plane ones $\eta_{\perp}$
   (right panel) as a function of $J^{\prime}$ for several lines
   with $J_{\perp}=$const.}
   \label{figi}
\end{figure*}
\section{Quantum phase diagram}\label{gen}
Quantum fluctuations radically modify the classical picture. Long-range
correlations with constant values of correlation functions, independent
on distance, are substituted by either quasi-long-range ones with
power-law behavior at large distances, or by short-range
exponentially-decaying ones.
For example, recently, it has been shown that quantum fluctuations
appreciably reduce the ordering amplitude in the chiral phase of the
model~(\ref{ham}) supplied with the Dzyaloshinskii-Moriya interaction
term~\cite{furukawa_00}. 
In the quantum phase diagram of~(\ref{ham})
we observe five phases as shown in the right panel of Fig.\ref{fig_phd}.

In our previous studies~\cite{Plekhanov_01} we have located the position
of one of these phases (Ferro) within the range $0<J_{\perp}<J_z$ and
$J^{\prime}\lesssim 0.31 J_z$. Contrarily to the isotropic Heisenberg
ferromagnet, where the ground state is $L+1$-times degenerate, in the
anisotropic model~(\ref{ham}) the ferromagnetic phase has only two
degenerate ground states with all spins either ``up'' or ``down''.
Therefore, the Ferro phase is structurally identical to that of the
classical phase diagram, the only difference being the shape of the
phase boundary.

We now describe the remaining four phases of
quantum phase diagram in Fig.~\ref{fig_phd}, which we have found
by use of the methods reported in Section~\ref{model}. 
\subsection{Spin-fluid-I ($XXZ$-like) phase}\label{xylike}
Above the ferromagnetic phase ($J_{\perp}>J_z$) and for moderate values
of $J^{\prime}$ (interpolating linearly the phase
boundary between the Spin-Fluid-I and E-II phases in Fig.~\ref{fig_phd}b,
we can write for the boundary $J_{\perp} \sim 2.6 J^{\prime} + 0.35J_z$)
we find a phase that we called Spin-fluid-I or $XXZ$-like because of
its similarity to the ground state of AF $XXZ$ model. A typical behavior
of the correlations in this phase is shown in
Fig.~\ref{fig_xxz}. The relevant in-plane correlations
are antiferromagnetic and power-law decaying, with periodicity vector
$\pi$ and critical exponent $\eta_{\perp}$, which is a function of both
$J_{\perp}$ and $J^{\prime}$:
\[
   \langle S^{x}_{r} S^{x}_{r+d} \rangle \sim (-1)^d d^{\eta_{\perp}}.
\]
For what concerns the out-of-plane channel, the relevant
correlations are always ferromagnetic:
\[
   \langle S^{z}_{r} S^{z}_{r+d} \rangle \sim -d^{\eta_{z}},
\]
with the exponent close to $-2$.
This fact confirms once again the analogy with $XXZ$ model.

Since DMRG deals with open boundary conditions (OBC), it is important to
keep the boundary effects under control. With OBC, a two-site
correlation function depends not only on the distance between the two
sites, but also on the position of these sites. To minimize boundary
effects, in the measurements of the two-site correlations, we choose
these two sites as symmetric as possible with respect to the center of
the cluster. In this way, the boundary effects, owing to the
nonequivalence of the lattice points under OBC, can be overcome. We
have studied the dependence of our results on the cluster size for a
typical point in the Spin-fluid-I phase. As shown in
Fig.~\ref{fig_fss1}, a progressive increase of $L$ increases the portion
of points, lying on a straight line, common for all values of $L$ under
investigation. The greater is the absolute value of
$\eta_{\perp}$($\eta_z$), the smaller are the boundary effects, since
for large a exponent the correlations decay fast enough on the scale of
the cluster size $L$.

The dependence of $\eta_{\perp}$ on coupling constants along several
lines with fixed values of $J_{\perp}$ and $J^{\prime}$ are shown in
Fig.~\ref{fig_xx}, left and right panels respectively. $\eta$ appears to
be monotonically decreasing as a function of $J^{\prime}$. In the case
when $J_{\perp}$ is constant, upon approaching the transition towards
the E-II phase (see Fig.~\ref{fig_phd}), $\eta_{\perp}$ shows a
tendency to diverge to minus infinity. Peculiarly, all the curves cross
at the same point around $J^{\prime}\approx 0.3J_z$, which means that at
this point $\eta_{\perp}$ is independent of $J_{\perp}$.

In order to check the quality of our DMRG data, we compare our results
for $\eta_{\perp}$ at the line $J^{\prime}=0$ with the known ``exact''
result of bosonization for the $XXZ$ model given by the
formula~(\ref{xy-1}). One can see from the right panel of
Fig.\ref{fig_xx} that our points fall exactly on the bosonization line,
meaning that our calculations (DMRG and the chosen cluster size) can
access the bulk low-energy physics, described by the bosonization.
Between $J^{\prime}=0.27J_z$ and $J^{\prime}=0.3J_z$ there is a change
of convexity of $\eta_{\perp}(J_\perp)$. At $J^{\prime}\approx
0.28J_z$, $\eta_{\perp}(J_{\perp})$ is approximately independent on
$J_{\perp}$, especially for $J_{\perp}$ far from $1$, confirming the
presence of a crossing point. The limiting value at $J_{\perp}\to\infty$
of the critical exponent for the in-plane correlations appears to be
$-1/2$, as follows also from~(\ref{xy-1}). This means that for
$J_{\perp}$ large enough $J^{\prime}$ becomes always irrelevant.
\subsection{Spin-fluid-II phase}\label{inc_i}
We identify another spin-fluid phase in the range $J^{\prime}>0.43J_z$,
$J_{\perp}< 1.8 J^{\prime}-0.2J_z$ {\it i.e.} when $J^{\prime}$ becomes
dominant, while the NN interactions (parameterized by $J_{\perp}$ and
$J_z$) become marginal. In Fig.~\ref{fig_ii} we report a typical
correlation picture in the Spin-fluid-II phase, while the raw data for
both in-plane and out-of-plane correlation functions are shown in the
respective insets. It is more convenient to plot the correlation
functions separating the inter- and intra-chain parts. Both in- and
out-of-plane correlation functions have antiferromagnetic character in
the intra- and inter-chain channels with periodicity vector
$q\sim\pi/2$. The power-law decay of the
correlations becomes clear after applying the logarithmic scale on both
axes. Although the dominant NNN interaction ($J^{\prime}$) is isotropic,
the corrections coming from the anisotropic NN ones ($J_{z}$, and
$J_{\perp}$) induce anisotropy in the correlations.

The out-of-plane correlations appear to have the same exponent $\eta_z$
for both intra- and inter-chain correlations. Moreover, for the
intra-chain correlations, there is an additional modulation by a
harmonic term $\sim(1+\alpha\sin(2\pi x/4))$, where $x$ is the distance
along a leg of the ladder and $\alpha\ll 1$. Such modulation, might be
a trace of a less-relevant correlations ({\it i.e.} of a correlation
with smaller exponent) which we are not able to measure directly.
Contrarily to the out-of-plane channel, in the in-plane one the
exponents of intra- ($\eta_{\perp1}$) and inter-chain ($\eta_{\perp2}$)
correlations appear to be slightly different, as shown in
Fig.~\ref{fig_ii}. The cluster-size dependence of the intra-chain
correlations is shown in Fig.~\ref{fig_fss2}. Once again, by increasing
the system size $L$, the power-law character of the correlations becomes
more and more evident. The convergence to the thermodynamic limit,
however, is somewhat slower in comparison with the Spin-Fluid-I phase
(see Fig.\ref{fig_fss2}), since now we are measuring the correlations
along a leg of the ladder and the maximum distance along a leg is only
half of the system size.

The critical exponents for the intra- and inter-chain correlations
are not universal and depend on the particular values of $J^{\prime}$
and $J_{\perp}$. $J^{\prime}$ dependence of the intra-chain correlation
exponents for in-plane $\eta_{\perp}$ and out-of-plane $\eta_z$ is shown
in Fig.~\ref{figi}. For $J_{\perp}<0.9, J_z$ $\eta_z<-1$ and
$\eta_{\perp}>-1$, while for $J_{\perp}>0.9 J_z$ $\eta_z>-1$ and
$\eta_{\perp}<-1$. When $J^{\prime}$ increases, both exponents tend to
$-1$ as in this case the system resembles more and more the $XXZ$ model
on two noninteracting legs.
\subsection{E-I and E-II phases}\label{mass}
The two phases E-I and E-II are characterized by the presence of at
least one exponentially-decaying correlation function. Actually, the
issue whether a gap is present in the system's spectrum is subtle and
would require a comprehensive study of all possible excitations.

The two phases (E-I, E-II) exhibit qualitatively different
correlations and the transition line between them goes from the point
($J^{\prime}=J_z/4$, $J_{\perp}=J_z$) to the point
($J^{\prime}=0.41J_z$, $J_{\perp}=0.5J_z$), as shown in
Fig.~\ref{fig_phd}. The transition lines are determined by the behavior
of the correlation functions. In a massless mode, the correlation
functions are linear in the $\log$-$\log$ scale (power-law decay), while
for a massive mode they are linear in the semi-$\log$ scale
(exponential decay).

\begin{figure}
   \includegraphics[angle=270, scale=0.33]{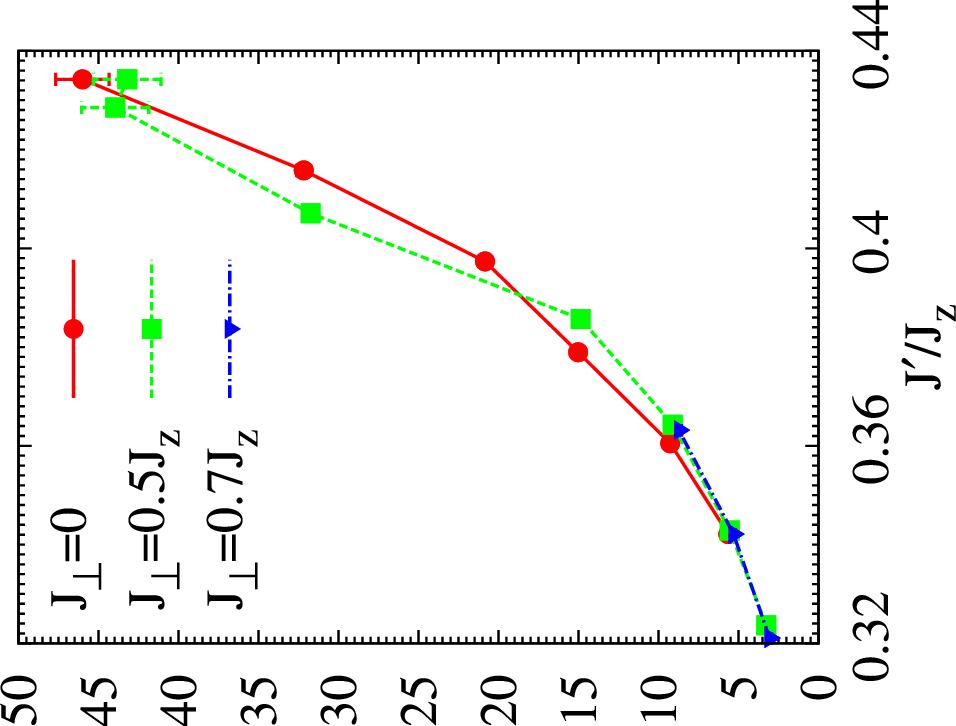}
   \hspace{0.4cm}
   \includegraphics[angle=270, scale=0.34]{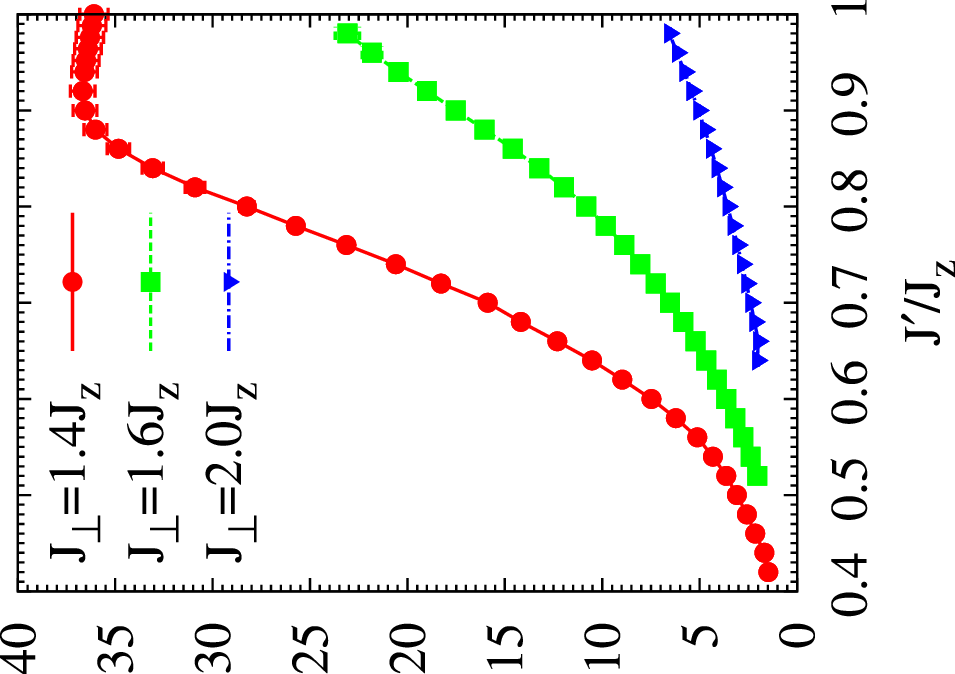}
   \caption{(Color online). $300$ site cluster. $J^{\prime}$ dependence
   of the correlation lengths for the in-plane correlations in the E-I
   phase (left panel) and for the out-of-plane ones in the E-II phase
   (right panel) for several values of $J_{\perp}$. Shown in this
   panel are the correlations along the legs ($d_0$ and $d_2$
   sequences as explained in the text). Lines are guides to the eye.
   }
   \label{fig3}
\end{figure}
\begin{figure*}
   \includegraphics[angle=270, width=0.47\textwidth]{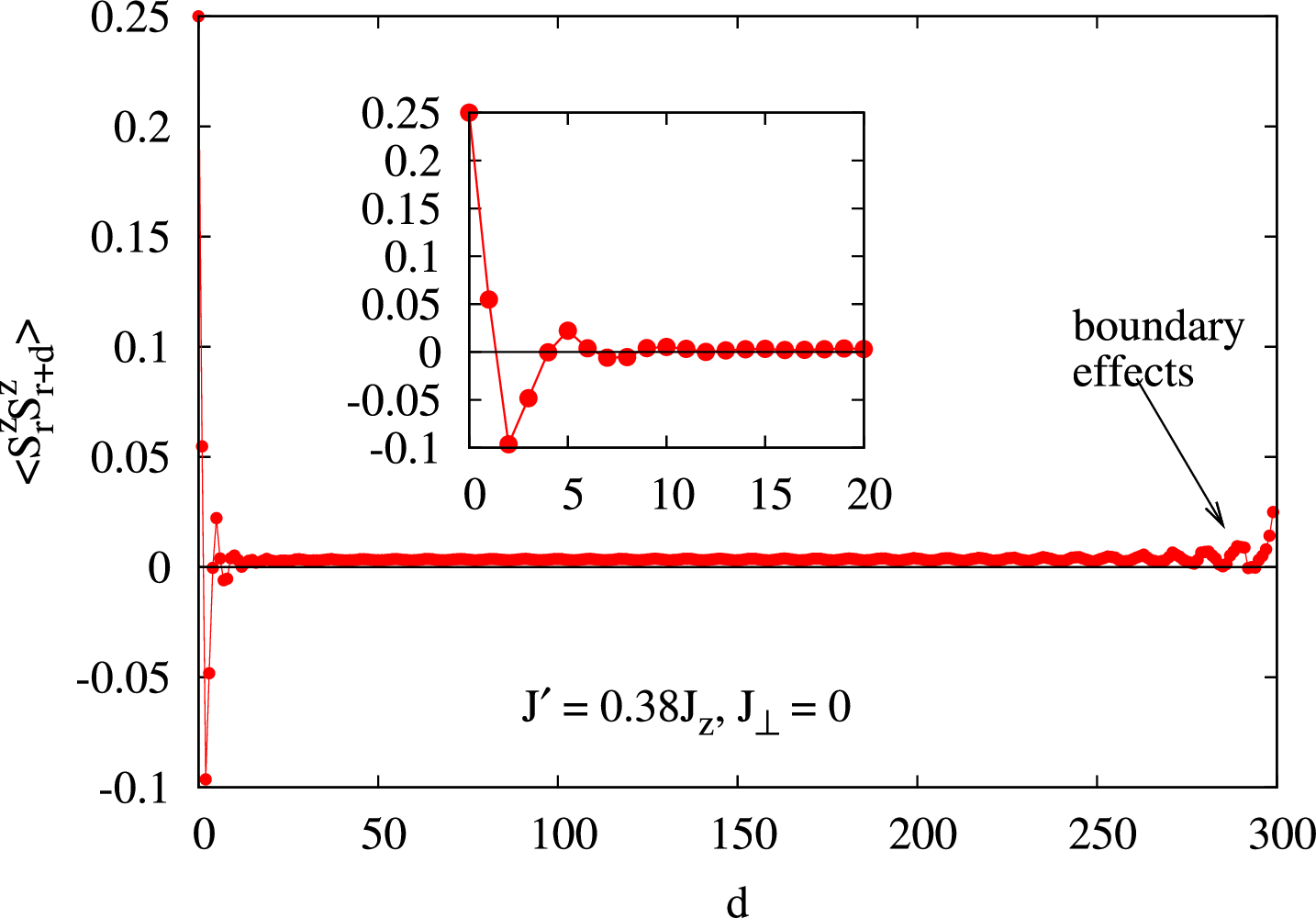}
   \hspace{0.4cm}
   \includegraphics[angle=270, width=0.47\textwidth]{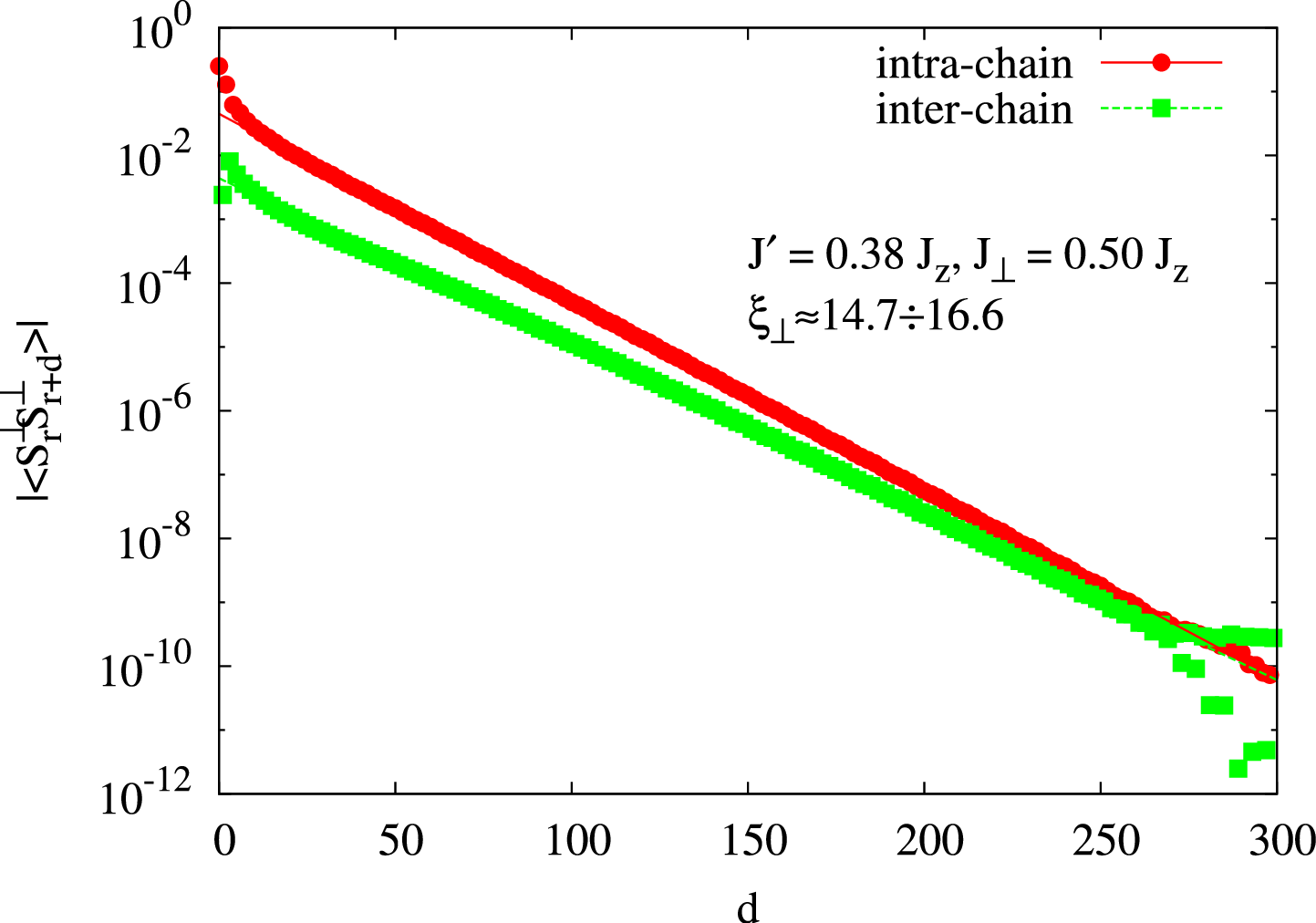}
   \caption{(Color online). $300$ site cluster. A sample of the
   correlations behavior in the E-I phase. Left panel: distance
   dependence of the out-of-plane static spin form factor.
   Short-range part of the correlations is shown
   in the inset. The shift due to finite magnetization $m=0.06$ is equal
   to $m^2=0.0036$ for the given values of $J^{\prime}$ and $J_{\perp}$.
   Right panel: semi-log plot of the in-plane spin form-factor. Intra-
   and inter-chain correlations have slightly different correlation
   lengths. Boundary effects manifest starting from approx. $d\gtrsim
   250$. Lines are guides to the eye.
   }
   \label{fig2}
\end{figure*}
The first phase (E-I) is located in the region
$0.31J_z<J^{\prime}<0.43 J_z$, $0<J_{\perp}<J_z$. E-I is adjacent to the
ferromagnetic phase and shows finite magnetization in the vicinity of
the transition line, which gradually goes to zero for $J^{\prime}
\gtrsim 0.35 J_z$. In the E-I phase, residual ferromagnetic correlations
induce a very complex structure of the static spin form factor in the
$z$-direction. In addition to the residual finite magnetization, which
shifts up the whole plot of $\langle S^{z}_{r} S^{z}_{r+d} \rangle$, the
boundary effects are enhanced with respect to what we find in the other
phases and in the in-plane channel. That is why we explore this phase on
clusters of the maximal reachable size on our machines: $300$ sites. At
short distances ($d<20$), in the out-of-plane channel the correlations
are dominated by an exponential incommensurate contribution, as shown in
the left panel of Fig.~\ref{fig2}. Qualitatively different behavior
appears for $d>20$. At medium distances ($20<d<250$) we observe
incommensurate long-range correlations with apparently no decay.
Finally, when the spins are situated at the opposite sides of the
cluster, the correlations between them grow as the spins approach the
borders of the cluster, being this behavior a clear evidence of a
boundary effect.

\begin{figure*}
   \includegraphics[angle=270, scale=0.38]{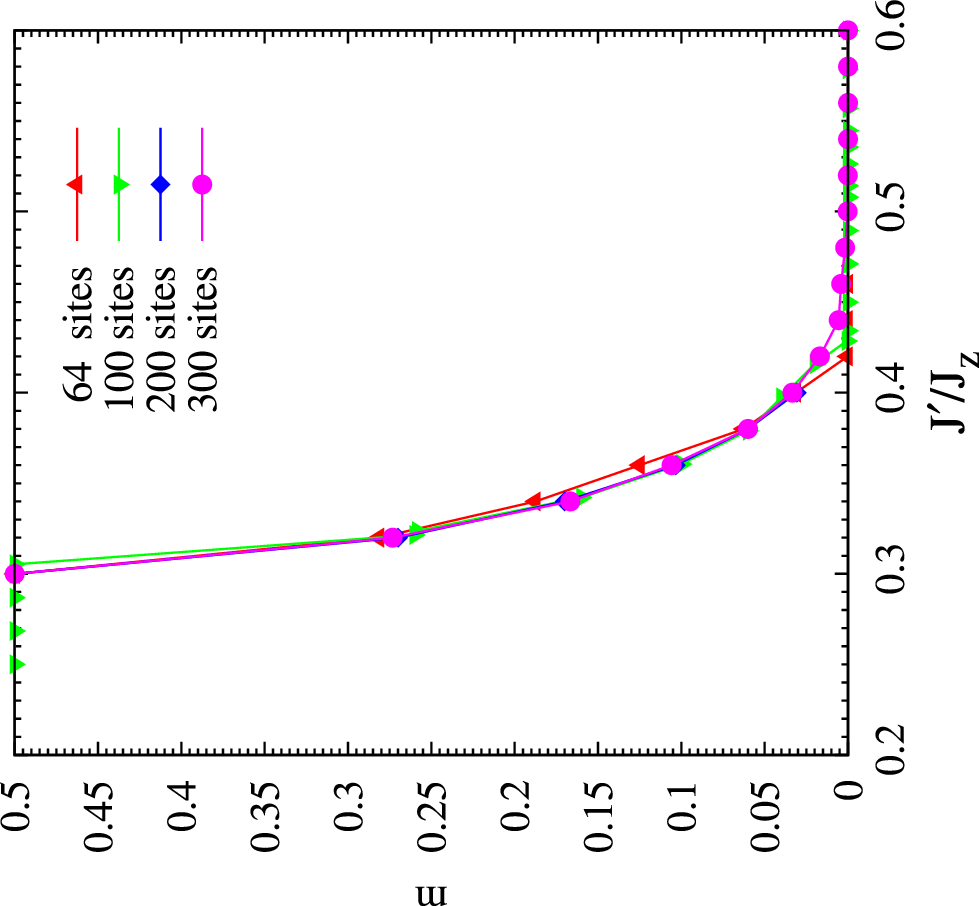}
   \includegraphics[angle=270, scale=0.38]{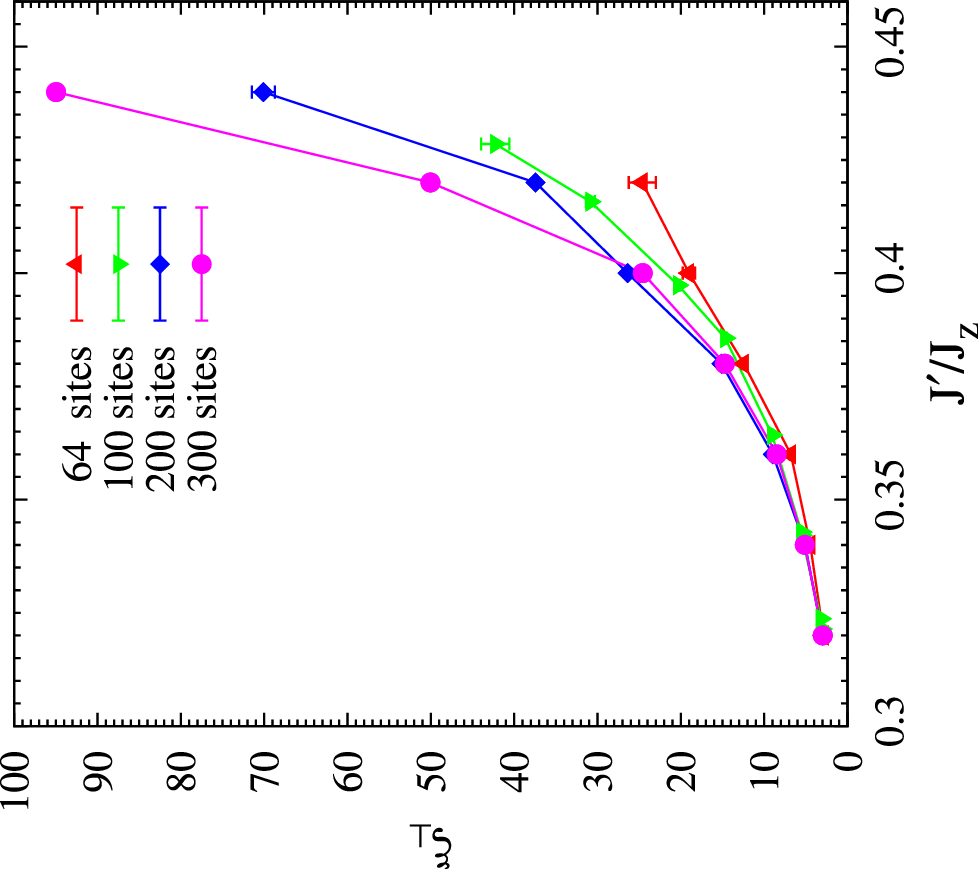}
   \hspace{0.1cm}
   \includegraphics[angle=270, scale=0.38]{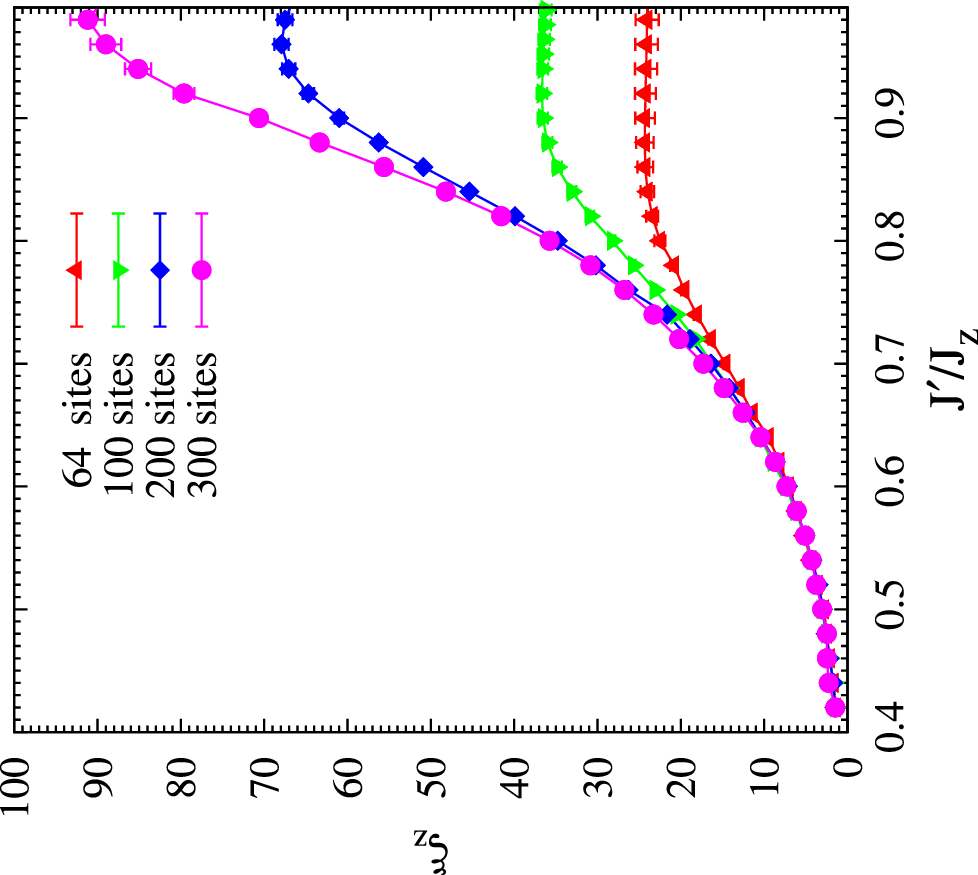}
   \caption{(Color online). $m$ - magnetization per site (left panel)
   and $\xi_{\perp}$ - correlation length of the in-plane correlations
   (middle panel) as a function of $J^{\prime}$ at $J_\perp=0.5J_z$ in
   the E-I phase for different system sizes. Right panel: $\xi_z$ - correlation
   length of the out-of-plane correlations in the E-II phase as a
   function of $J^{\prime}$ for different system sizes.
   }
   \label{fig5}
\end{figure*}
\begin{figure*}
   \includegraphics[angle=270, width=0.47\textwidth]{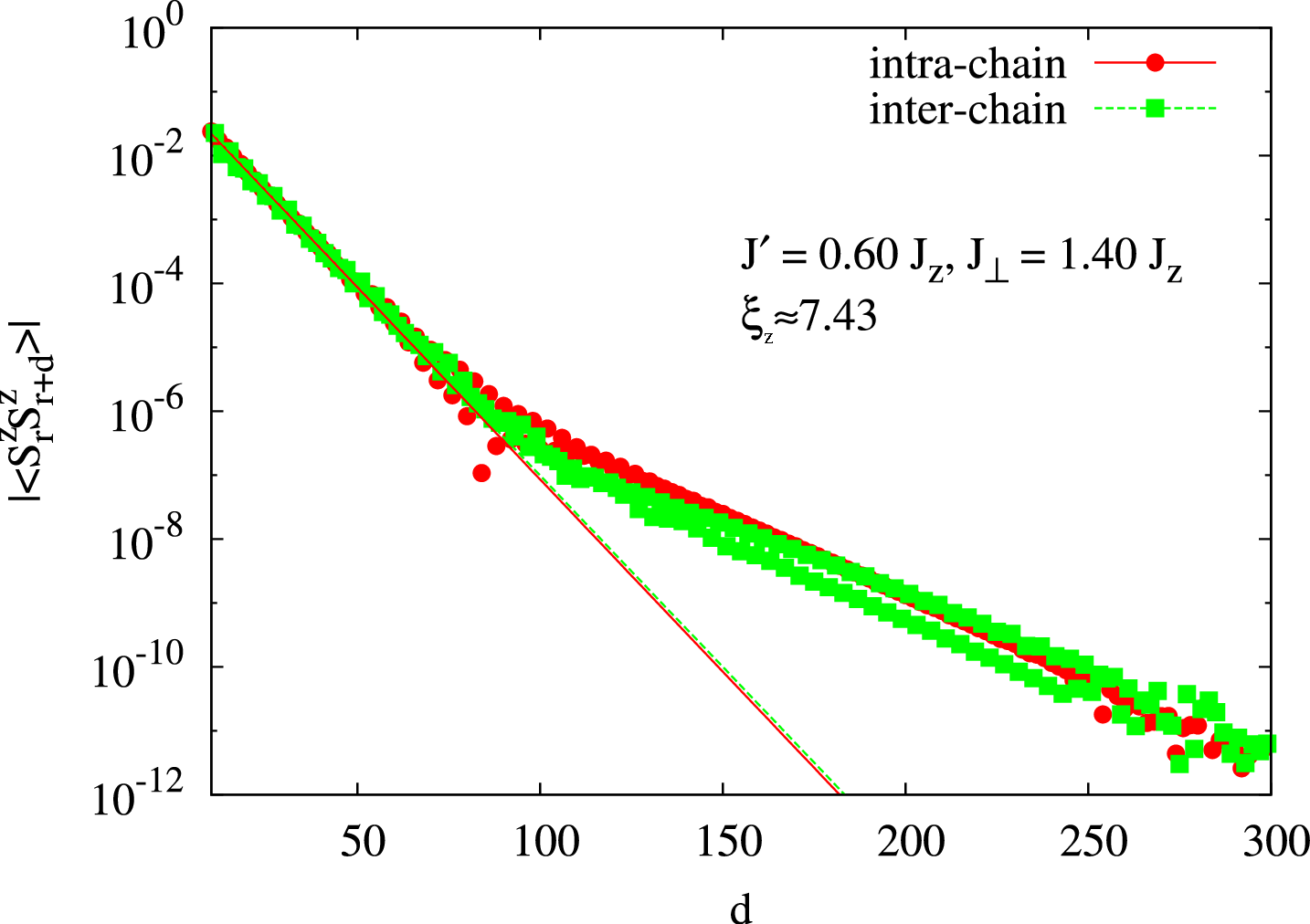}
   \hspace{0.2cm}
   \includegraphics[angle=270, width=0.47\textwidth]{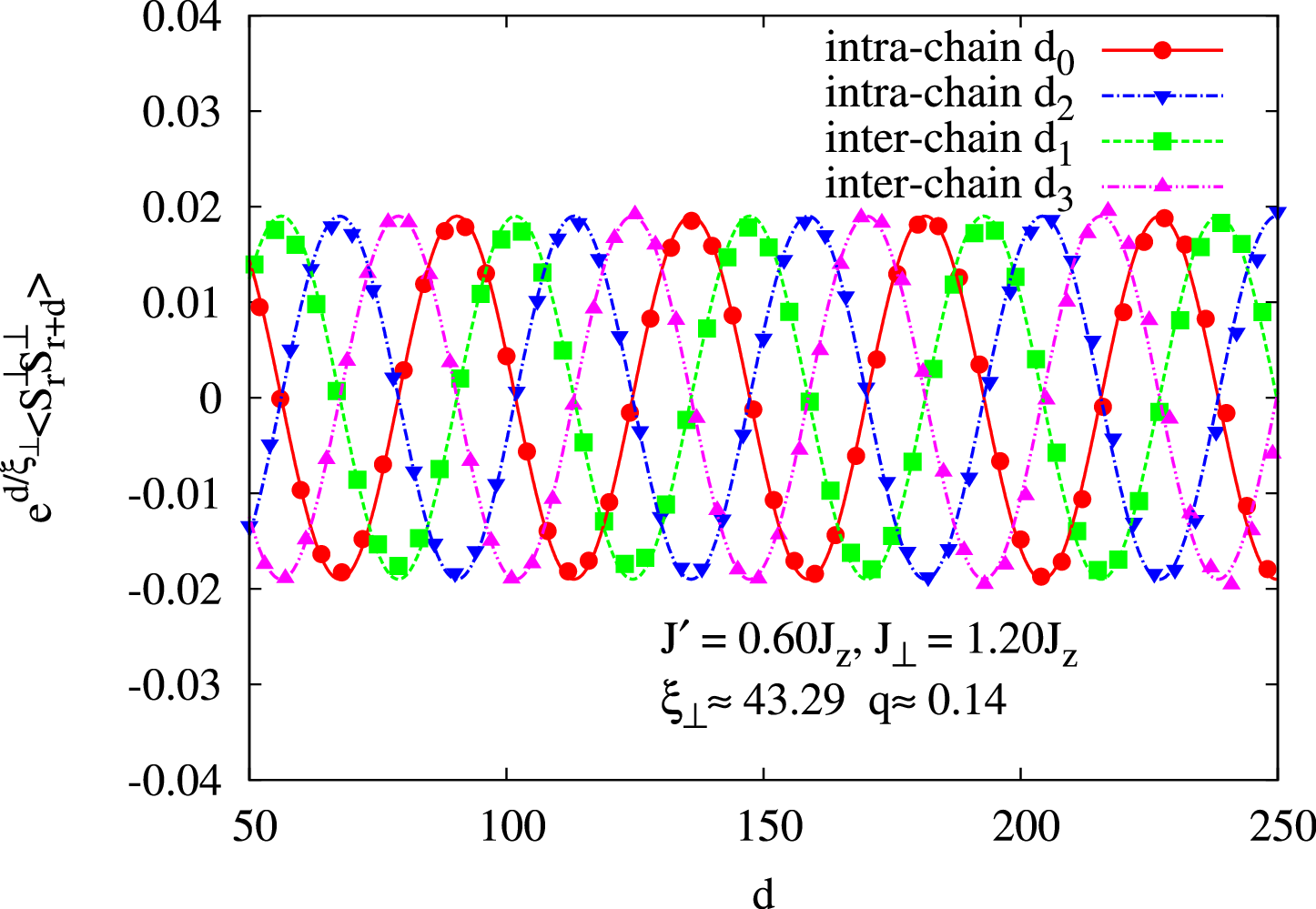}
   \caption{(Color online).
   $300$ site cluster. A sample of the correlations behavior in the E-II
   phase. Left panel: semi-log plot of the out-of-plane spin
   form-factor. Both intra- and inter-chain correlations fit the same
   line. For $d\gtrsim 100$, $|\langle S^{z}_{r} S^{z}_{r+d}
   \rangle|\lesssim 10^{-7}$, which is less than the accuracy of our DMRG
   method and such data are reported solely for the sake of completeness
   and reproducibility.
   Right panel: in-plane spin form-factor multiplied by the
   exponential pre-factor $e^{d/\xi}$. The points sequences fit quite
   well on the cosine functions with different phase shifts and with the
   incommensurate periodicity vector $q=0.14$. Only distances where the
   exponential-decaying behavior is established are shown. Lines are
   guides to the eye.
   }
   \label{fig4}
\end{figure*}

As it can be seen from the right panel of Fig.~\ref{fig2}, the in-plane
correlations exhibit exponential decay. As we have already mentioned,
the model~(\ref{ham}) can be also considered as a zig-zag ladder. Hence,
there could be an \textit{a priori} distinction between intra- and
inter-chain correlations. The correlations we measured distinguish
whether the two sites stay on the same leg or not. The inter-chain
in-plane correlations have smaller correlation length with respect to
the intra-chain ones, and go gradually to zero as $J_{\perp}\to0$, since
in that case the two legs of the ladder remain coupled only in the
$z$-channel. In the left panel of Fig.~\ref{fig3}, we plot the
$J^{\prime}$ dependence of the in-plane intra-chain correlation length
$\xi_{\perp}$ for several values of $J_{\perp}$.
$\xi_{\perp}(J^{\prime})$ appears to be independent of $J_{\perp}$ at
least close to the transition line Ferro$\to$E-I.
The maximal value reached by $\xi_{\perp}$ ($50$ lattice
constants) on approaching the transition E-I$\to$Spin-fluid-II (along
$J_{\perp}=0.0$ and $0.5J_z$ lines) is much greater than that
($10$ lattice constants) reached at the transition E-I$\to$E-II (along
$J_{\perp}=0.7J_z$ line).

The issue of finite magnetization in the E-I phase, rather unexpected,
deserves a separate study through the finite-size scaling in order to
conclude whether it might be a bulk feature.

In order to determine the residual magnetization, we have used the DMRG
program with symmetry reduction of the Hilbert space into invariant
subspaces with definite $S^{z}_{tot}$. In such a case, the residual
magnetization at given values of $J_{\perp}$ and $J^{\prime}$
corresponds to the sector of $S^{z}_{tot}$ containing the global energy
minimum. These results were subsequently verified by another DMRG
program without symmetry reduction where at each point of the phase E-I
the absolute value of the residual magnetization has been deduced from
the sum of the $z-$component of spin-spin correlation functions on for all
distances:
\be
   |m| = \frac{ |\langle S^{z}_{tot} \rangle| }{L} = 
   \frac{1}{L}\sqrt{\sum_{i,j} \langle S^{z}_{i} S^{z}_{j}\rangle}.
   \label{magn}
\ee
In~(\ref{magn}) we have used the fact that in absence of the magnetic
field and with rotational degeneracy broken $J_{\perp}\neq J_z$, the
eigenstate with a $S^{z}_{tot}=\mathcal{M}$ is degenerate with respect to
$S^{z}_{tot}=-\mathcal{M}$. Therefore, for the program without symmetry reduction
a ground state with $S^{z}_{tot}=|\mathcal{M}|$ will be an arbitrary mixture of
states with $S^{z}_{tot}=\mathcal{M}$ and $S^{z}_{tot}=-\mathcal{M}$, but $|S^{z}_{tot}|$
will have a definite value.
\newc{
In order to exclude that the residual magnetization in the E-I phase is
a DMRG artifact caused by the difficulties to reach convergence in the
proximity of a phase transition, we have performed a finite-size scaling
analysis by means of Lanczos technique, free of DMRG truncation error,
on clusters with up to 30 sites for both open and periodic boundary
conditions. Lanczos data has confirmed DMRG ones: in the E-I phase, the
finite magnetization has no tendency to disappear also in the
thermodynamic limit.
}

In the E-I phase and
for the values of $L$ accessible to us, we find a close relation between
the finite residual magnetization and the correlation length
$\xi_{\perp}$ in the in-plane channel, as shown in the left and middle
panels of Fig.~\ref{fig5}. Namely, the decrease of the magnetization per
site is accompanied by the increase of $\xi_{\perp}$. The former appears
to be almost size independent for $J^{\prime}<0.4J_z$. At
$J^{\prime}>0.4J_z$, finite magnetization steps of height $1/L$
appear. This occurs because, for a finite cluster, the $z$-component of
the total spin can only increment by one. The disappearance of
magnetization is accompanied by the divergence of the correlation length
$\xi_\perp$ in the in-plane channel (left and middle panels of
Fig.~\ref{fig5}). On increasing $J^{\prime}$, $\xi_\perp$ reaches a
maximal value (of the order of $L/2$) at $J^{\prime}=0.42\div 0.44J_z$
and, for greater values of $J^{\prime}$, a power-law fit appears to be
more appropriate.

Such behavior is surprisingly similar to what was found in the
multi-magnon phase of the isotropic model~(\ref{ham}) at
$J^{\prime}>\frac{1}{4}J_z$ and in magnetic
field~\cite{drechsler_04,kecke_01,kecke_02}. Indeed, a state with $n$
magnons is characterized by a wave function with non-zero total
magnetization equal to $\pm(L/2-n)$. It is worth noting that for
$J^{\prime}>\frac{1}{4}J_z$ finite magnetization can exist in the
isotropic model only in the presence of an external magnetic field,
while in our case it can be induced by tuning the values of $J^{\prime}$
and $J_{\perp}$ for zero field. Increasing $J^{\prime}$ at fixed
$J_{\perp}$ the total cluster magnetization decreases from $L/2$ to
zero, implying that the total number of magnons increases from zero to
$L$ (infinity in thermodynamic limit). In the case of the isotropic
model in magnetic field, the low-lying excitations are generally formed
of bound multi-magnon clouds and are gapless, while the in-plane
correlation functions are expected to decay
exponentially~\cite{kecke_01,kecke_02}. If the multi-magnon scenario is
realized also in our case is currently under investigation. As a matter
of fact, the exponential decay of the in-plane correlations (see the
left panel of Fig.\ref{fig2}) is clearly in favor of this scenario.

In the E-II phase, when both $J^{\prime}$ and $J_{\perp}$ become large
enough, the correlations change qualitatively. In this phase, the ratio
$J_{\perp}/J^{\prime}$ is limited from below and above. The out-of-plane
correlations exhibit an exponential decay as functions of the distance
between any two sites, regardless whether they belong to the same or
different legs (see left panel of Fig.~\ref{fig4}). We have also
investigated $J^{\prime}$ dependence of the out-of-plane intra-chain
correlation length $\xi_z$ for several values of $J_{\perp}$ in the E-II
phase, as reported in the right panel of Fig.~\ref{fig3}.
$\xi_z(J^{\prime})$ grows monotonically from the left border of the
phase (transition Spin-fluid-I$\to$E-II) to the right one (transition
E-II$\to$Spin-fluid-II). Such non-symmetric behavior implies radically
different types of orders in the two massless phases.

The in-plane correlation function shows an exponential decay with a
correlation length $\xi_{\perp}$ ($\xi_{\perp}=43.29$ for $J^{\prime}=0.6
J_z$ and $J_{\perp}=1.2J_z$). Once multiplied by $e^{d/\xi_{\perp}}$, as
shown in the right panel of Fig.~\ref{fig4}, the correlation function
presents incommensurate oscillations as a function of $d$ with
periodicity vector $q$ ($q=0.14$ for $J^{\prime}=0.6 J_z$ and
$J_{\perp}=1.2J_z$) and a phase depending on ${\rm mod}(d,4)$ as
follows. The whole set of distances $d\equiv|i-j|$ between any two sites
$i$ and $j$ on a 1D cluster splits into four sequences:
\begin{eqnarray}
   \nonumber
   &&d_0(l) = 4 l + 0, \qquad 0,4,8,12,\dots \\
   \nonumber
   &&d_1(l) = 4 l + 1, \qquad 1,5,9,13,\ldots\\
   \nonumber
   &&d_2(l) = 4 l + 2, \qquad 2,6,10,14,\ldots \\
   \nonumber
   &&d_3(l) = 4 l + 3, \qquad 3,7,11,15,\ldots
\end{eqnarray}
where $l=0,\ldots, 24$ for a $300$ site cluster. One can easily verify by
looking at Fig.~\ref{fig_m} that the sequence $d_0(l)$ corresponds to the
case when both $i$ and $j$ belong to the same leg and $i-j$ is even.
Similarly, the sequence $d_2(l)$ comprises all the cases when both $i$
and $j$ belong to the same leg and $i-j$ is odd. The remaining two
sequences $d_1(l)$ and $d_3(l)$ are realized when $i$ and $j$ belong to
different legs. All four sequences appear to be modulated by the
same harmonic term, but with different phases. Peculiarly, the phase
shift between $d_0(l)$ and $d_2(l)$ equals to $\pi$ and the same does
the phase shift between $d_1(l)$ and $d_3(l)$. Finally, the phases of
$d_0(l)$ and $d_1(l)$ differ by $\pi/2$. Summarizing, we can conclude
that the in-plane correlations of the E-II phase assume the following
asymptotic form:
\begin{equation}
   \langle
   S^{x}_{n} S^{x}_{n+d}
   \rangle
   \sim \exp(-d/\xi_{\perp}) 
   \cos(qd+\phi_d),
\end{equation}
where $\phi_d=\frac{\pi}{2}{\rm mod}(d,4)$.

On a cluster of $100$ sites, the out-of-plane correlation length
$\xi_{z}$ approaches its saturation value of about $\xi^{c}_{z}\sim
35a$, while the system undergoes the phase transition towards the
Spin-Fluid-II phase, and hence $\xi^{c}_{z}/L\sim 0.35$ for $L=100$.
This value should be compared to $L/2=50$ since we are speaking about
the intra-chain correlations. In the right panel of Fig.\ref{fig5} we
report the $J^{\prime}$-dependence of $\xi_{z}$ for different system
sizes. The ratio $\xi^{c}_{z}/L$ ranges from $0.3$ for $L=300$ to
$0.375$ for $L=64$ while the ``uncertainty'' range, {\it i.e.} the width
of the transition on a finite system, decreases as expected.

The phase E-II comprises the isotropic line for the values
$J^{\prime}\in[0.25J_z,0.7J_z]$. In this region, an "astronomically
small" gap~\cite{itoi} has been predicted by means of bosonization. The
vanishing value of the predicted gap precludes any possibility of its
numerical observation. Indeed, in the E-II phase, finite-size scaling
shows that the excitation gaps, both in singlet sector (containing the ground state)
and between triplet and singlet sectors converge to zero in our DMRG
calculations (not shown).

Currently, by using the numerical tools to our disposition, we can not
discriminate between the two hypotheses: i) that there indeed exists a
tiny gap in the spectrum, which can not be directly observed
numerically, and ii) that the system is gapless, but the excitations
associated with the correlation functions studied have a gap. 
%
\section{conclusions}\label{concl}
We have performed an extensive numerical study of the model~(\ref{ham})
within the following range of the Hamiltonian parameters: $J_z=1$,
$0<J_{\perp}<2.5$, $0<J^{\prime}<1.0$. We have constructed a
zero-temperature phase diagram based on the DMRG calculations for
$64,100,200,300$ site system with open boundary conditions.

We have identified several phases within this phase diagram, based on
the behavior at large distances of the correlation functions. We have
found two phases with quasi-long-range behavior, among which one appears
to be a generalization of the ground state of $XXZ$ model, and another
possesses rather complex correlation picture, which distinguishes
between intra- and inter-chain distances. We have discovered two phases
with exponentially-decaying correlation functions. One of them (E-II)
separates the two massless phases, while the other (E-I) divides a
massless phase from the ferromagnetic one. Although these two phases
present qualitatively different correlations, they share a common
feature. Both of them develop when one of the NN couplings is less
efficient respect to the other and to $J^{\prime}$. Indeed, in the phase
E-I, $J_{\perp}$ is limited respect to the other couplings, while in
E-II the relative weight of $J_z$ goes to zero with respect to
$J_{\perp}$ and $J^{\prime}$.
A further insight into the nature
of the phases found in this manuscript can be obtained by studying more
specific properties of their ground states, like entanglement, or by
measuring various more sophisticated correlation functions {\it e.g.}
spiral or dimer ones~\cite{kaburagi_00,kaburagi_01,kaburagi_02}. Such
work is currently in progress.

For what concerns real materials, for ${\rm Rb_2Cu_2Mo_3O_{12}}$ we
have~\cite{hase} $J^{\prime}\approx0.37J_z$ and from Fig.~\ref{fig_phd}
we can conclude that its ground state should be either in E-I or E-II
phase depending on the anisotropy. The same conclusion holds also for
${\rm Li_2 ZrCuO_4}$ ($J^{\prime}\approx0.26J_z$), ${\rm Pb_2
[CuSO_4(OH)_2]}$ ($J^{\prime}\approx0.43J_z$) and ${\rm
Cs_2Cu_2Mo_3O_{12}}$ ($J^{\prime}\approx0.38J_z$) as estimated in the
Ref.~\cite{drechsler_02}. ${\rm NaCu_2O_2}$ was
found~\cite{drechsler_01} to have $J^{\prime}\approx2J_z$ and assuming
that the anisotropy is not too strong we conclude that its ground state
is in the Spin-Fluid-II phase. Finally, for ${\rm LiCuVO_4}$ the
estimates for $J^{\prime}$ range from $2.4J_z$ to $3.5J_z$. These
values, under the assumption of weak anisotropy, bring us again to the
Spin-fluid-II phase. Recently, a new value of $J^{\prime}\sim 0.3J_z$
for Li$_2$CuO$_2$ has been deduced by means of various
methods~\cite{drechsler_05,drechsler_06}, in contrast to the value
$J^{\prime}=0.62J_z$ previously known~\cite{mizuno_00}. If one
approximates Li$_2$CuO$_2$ by non-interacting chains each described by
Hamiltonian~(\ref{ham}), then this material should belong either to E-I
or E-II phase depending on the actual anisotropy (still under debate).
\begin{acknowledgement}
It is our greatest pleasure to acknowledge stimulating discussions with
A. Chubukov and S.-L Drechsler and A.A. Nersesyan. We wish to thank the
Referees for making several suggestions that substantially improved the
readability of the manuscript. 
\end{acknowledgement}
\bibliographystyle{epj}

%
\end{document}